\documentclass[showpacs,amsmath,amssymb,aps,preprint]{revtex4} 
\input{epsf}
\usepackage{graphicx}

\begin{document}


\title{Slowly Rotating General Relativistic Superfluid Neutron Stars 
with Relativistic Entrainment}

\author{G.~L.~Comer}
\email{comergl@slu.edu}
\affiliation{Department of Physics, Saint Louis University, St.~Louis, 
MO, 63156-0907, USA}

\date{\today}

\def\beq{\begin{equation}}
\def\eeq{\end{equation}}
\def\n{n}
\def\p{p}
\def\d{\delta}
\def\A{{\cal A}}
\def\B{{\cal B}}
\def\C{{\cal C}}
\def\a00{{{\cal A}_0^0}}
\def\b00{{{\cal B}_0^0}}
\def\c00{{{\cal C}_0^0}}
\def\D00{{{\cal D}_0^0}}
\def\M{{\cal M}}
\def\wh{\widehat}
\def\wa{\widetilde}
\def\ha{\overline}
\def\ms0{{\left.m_*\right|_{\rm o}}}

\begin{abstract}
Neutron stars that are cold enough should have two or more 
superfluids/supercondutors in their inner crusts and cores. \ The 
implication of superfluidity/superconductivity for equilibrium 
and dynamical neutron star states is that each individual particle 
species that forms a condensate must have its own, independent number 
density current and equation of motion that determines that current. \ 
An important consequence of the quasiparticle nature of each condensate 
is the so-called entrainment effect, i.e.~the momentum of a condensate 
is a linear combination of its own current and those of the other 
condensates. \ We present here the first fully relativistic modelling of 
slowly rotating superfluid neutron stars with entrainment that is 
accurate to the second-order in the rotation rates. \ The stars consist 
of superfluid neutrons, superconducting protons, and a highly 
degenerate, relativistic gas of electrons. \ We use a relativistic 
$\sigma$ - $\omega$ mean field model for the equation of state of the 
matter and the entrainment. \ We determine the effect of a relative 
rotation between the neutrons and protons on a star's total mass, shape, 
and Kepler, mass-shedding limit.
\end{abstract}


\maketitle

\section{Introduction}

There is now an extended body of evidence in support of superfluidity in 
dense, nucleonic matter, such as that which is believed to exist in 
neutron stars. \ Using conservative estimates one can argue for a Fermi 
temperature on the order of $10^{12}~{\rm K}$ for neutrons in media with 
supranuclear densities, and that the transition temperature to a 
superfluid state is about $10^9~{\rm K}$. \ This is a signficiant fact 
since it is generally accepted that nascent neutron stars formed from 
supernovae should cool fairly quickly, and consequently their internal 
temperatures should pass quickly through the transition value. \ 
Observational support for such transitions is supplied by the well 
established glitch phenomenon in pulsars \cite{RM69,L93}. \ These are 
rapid decreases in the rotational periods followed by a slow recovery 
\cite{RD69}, much too slow to be explained by ordinary fluid viscosity 
\cite{BPPR69}. \ The best description is based on superfluid quantized 
vortices, and how they pin, unpin and then repin as the pulsar's 
rotation rate evolves \cite{AI75,AAPS84a,AAPS84b}. \ We present here a 
fully relativistic formalism to model the rotational properties of 
superfluid neutron stars within a slow rotation approximation, which is 
an extension to second order in the rotation rates the previous work of 
Comer and Joynt \cite{CJ03}.

The bulk of the studies of neutron star superfluidity have been in the 
Newtonian regime, where the intent is not to be quantitatively 
descriptive but rather to gain qualitative understandings. \ However, it 
is becoming increasingly apparent that general relativity is required to 
obtain even qualitative understandings. \ And even if the qualitative 
does not vary between the two regimes, the number of examples is growing 
where general relativity can yield factors of two difference from 
Newtonian calculations (as opposed to ``merely'' $20\%$ to $30 \%$ 
corrections). \ For example, recent modelling of supernovae has revealed 
that when general relativity is included the range of model parameters 
that exhibit multiple bounces is significantly smaller than the range 
found for the Newtonian case \cite{DFM02}. \ The use of general 
relativistic hydrodynamics has also led to predictions of the shock 
radius (during the shock reheating phase) being reduced by a factor of 
two and a corresponding increase by a factor of two for the inflow speed 
of the material behind the shock \cite{BNM01}. \ Certainly the need to 
use general relativity must only be enhanced as supernova remnants 
become more compact.

Weber \cite{W99} provides an excellent overview of the many suggestions 
for the matter content of these remnants, the neutron stars. \ They 
range from the traditional neutron/proton/electron models to more exotic 
configurations with kaon or muon condensates in the core, or hyperons, 
or even strange stars with absolutely stable u, d, s quark matter. \ In 
our study we will be somewhat conservative by considering typical 
traditional neutron star models, composed of superfluid neutrons, 
superconducting protons (with proton fractions on the order of $10\%$), 
and a highly degenerate relativistic gas of electrons. \ Each species 
extends from the center of the star all the way to the surface, although 
in principle one could consider a more realistic scenario wherein the 
protons and electrons extend out further than the neutrons 
\cite{ACL02,PCA02,PNC02} and in that way mimic features of a crust. \ 
This same technique could be applied in neutron star cores, allowing 
for the possibility of alternating regions of ordinary fluid and 
superfluid.  

The local thermodynamic equilibrium of the matter will be modelled using 
a relativistic $\sigma$ - $\omega$ mean field approach of the type that 
is attributed to Walecka \cite{W95} and discussed in detail for neutron 
stars by Glendenning \cite{G97}. \ We consider a relativistic approach 
to be important on two different levels. \ On the macroscopic level 
there is the need for general relativity that was discussed above. \ But 
on a microscopic level, recall that any fluid approximation for matter 
has built into it the notion of local, fluid elements. \ They are small 
enough that they can be considered to be points with respect to the rest 
of the star, and yet large enough to contain, say, an Avogadro's number 
of particles. \ At the densities expected for neutron stars, the local, 
Fermi levels for the nucleons can become high enough that the effective 
velocities of the nucleons with respect to their fluid elements are 
relativistic. \ As for the fluid elements themselves, they can also, in 
principle, approach speeds near that of light, although in practice 
(e.g.~for quasinormal mode \cite{ACL02,CLL99} or slow rotation 
calculations \cite{AC01c,CJ03}) they will typically have speeds a few 
percent of that of light.

Because of the superfluidity of the neutrons, and superconductivity of 
the protons, the fluid formalism to be used differs fundamentally from 
the standard perfect fluid approach in that the neutrons can flow 
independently of the protons and electrons. \ There are thus two fluid 
degrees of freedom in the system, requiring two sets of fluid elements, 
one set for the neutrons and another for the charged constituents since 
the electromagnetic interaction very effectively ``ties'' the electrons 
to the protons. \ The matter description, therefore, must take into 
account two Fermi levels for the nucleons, and a displacement in 
momentum space between their respective Fermi spheres that will result 
when one fluid flows with respect to the other. \ The specification of a 
local thermodynamic equilibrium for the two fluids requires that the 
local neutron and ``proton'' (i.e.~a conglomeration of the protons and 
electrons) number densities be known as well as the local, relative 
velocity of the proton fluid elements, say, with respect to those of the 
neutrons. 

When the Fermi spheres for the nucleons are displaced with respect to 
each other there results an important effect for neutron star dynamics 
known as entrainment \cite{AB76}. \ Sauls \cite{S89} describes the 
entrainment effect using quasiparticle language, i.e.~just because the 
neutrons are superfluid and protons superconducting does not mean they 
no longer feel the strong force. \ On the contrary, an individual 
neutron should be understood as being surrounded by a polarization cloud 
of other neutrons and protons. \ When this neutron moves, it will be 
accompanied by this cloud of nucleons. \ The net effect at the level of 
the fluid elements is that the momentum of a neutron fluid element 
is a linear combination of the neutron {\em and} proton number density 
currents. \ Parameters that are important for entrainment in neutron 
stars have been calculated in the Newtonian regime using a Fermi-liquid 
approach \cite{BJK96} and in the relativistic regime via a $\sigma$ - 
$\omega$ mean field model \cite{CJ03}. \ Comer and Joynt \cite{CJ03} 
find that one of the key parameters that has been much used in 
superfluid neutron star modelling extends over a much larger range of 
values than what the Newtonian analysis of Borumand et al would imply 
\cite{BJK96}.  

Comer and Joynt also used their formalism to obtain first order 
rotational corrections to a superfluid neutron star's equilibrium state, 
which for general relativity means determining the frame-dragging, and 
the angular momentum. \ We will extend those calculations to the second 
order in the rotation rates, and thereby determine rotational 
corrections to the metric, distribution of particles, the total mass, 
shape, and the Kepler, mass-shedding limit. \ We construct sequences of 
equilibrium configurations where members of the sequence have the same 
relative rotation between the neutrons and protons but are distinguished 
by their central neutron number densities. \ We also construct sequences 
where the central neutron number density is fixed and the relative 
rotation is allowed to vary (cf.~\cite{AC01c,PCA02} for discussions on 
when to expect a relative rotation). \ One other straightforward, but 
important, extension here of the work of Comer and Joynt is a proof that 
the matter coefficients obtained by them are sufficient for our second 
order calculations.

In order to have a reasonably self-contained document, and to define all 
the variables, we review in Sec.~\ref{review} the general relativistic 
superfluid formalism and its application to slowly rotating neutron 
stars. \ It is in this section that we prove the matter coefficients 
obtained by Comer and Joynt are all that is needed for the extension to 
second order. \ In Sec.~\ref{meanfield} we discuss the highlights of the 
relativistic $\sigma$ - $\omega$ model and its mean field limit. \ We 
also determine the model's slow rotation limit. \ In Sec.~\ref{results} 
we join the slow rotation formalism with the mean field  model and  
produce numerical solutions. \ After reviewing the main results, the 
final, concluding section discusses applications beyond those considered 
here and points out where the formalism should be improved. \ For 
convenience we have restated in the appendix results of Comer and Joynt 
for the various matter coefficients that are required input for the 
field equations. \ We use ``MTW'' \cite{MTW} conventions throughout and 
geometrical units.

\section{General Relativistic Superfluid Formalism and Slow Rotation} 
\label{review} 

\subsection{The full formalism}

We will use the formalism developed by Carter, Langlois, and their 
various collaborators 
\cite{C89,CL94,CL95,CL98a,CL98b,LSC98,CLL99,P00,AC01c,C02}. \ The 
fundamental fluid variables consist of the conserved neutron $n^{\mu}$ 
and proton $p^{\mu}$ number density currents, from which are formed the 
three scalars $\n^2 = - \n_\mu \n^\mu$, $\p^2 = - \p_\mu \p^\mu$, and 
$x^2 = -p_\mu n^\mu$. \ Given a master function 
$- \Lambda(\n^2,\p^2,x^2)$ (i.e.~the superfluid analog of the equation 
of state), then the stress-energy tensor is  
\beq
    T^\mu_\nu = \Psi \delta^\mu_\nu + \n^\mu \mu_\nu + \p^\mu 
                \chi_\nu \ ,
\eeq
where 
\beq
    \Psi = \Lambda - n^\rho \mu_\rho - p^\rho \chi_\rho \label{press}
\eeq
is the generalized pressure and
\beq
    \mu_\nu = {\cal B} n_\nu + {\cal A} p_\nu  \quad , \quad
    \chi_\nu = {\cal A} n_\nu + {\cal C} p_\nu \ , 
\eeq
are the chemical potential covectors which also function as the 
respective momenta for the fluid elements. \ The $\A$, $\B$, and $\C$ 
coefficients are obtained from the master function via the partial 
derivatives
\beq
    {\cal A} = - \frac{\partial \Lambda}{\partial x^2} \ , \quad 
    {\cal B} = - 2 \frac{\partial \Lambda}{\partial \n^2} \ , \quad 
    {\cal C} = - 2 \frac{\partial \Lambda}{\partial \p^2} \ . \quad 
\eeq
The fact that the neutron momentum $\mu_{\mu}$, say, is not simply 
proportional to its number density current $\n^{\mu}$ is a result of 
entrainment between the neutrons and protons, which we see vanishes if 
the $\A$ coefficient is zero.

Finally the equations for the neutrons and protons consist of the two 
conservation equations 
\beq
    \nabla_{\mu} \n^{\mu} = 0 \quad , \quad 
    \nabla_{\mu} \p^{\mu} = 0 \ ,
\eeq
and the two Euler equations 
\beq
    \n^{\mu} \nabla_{[\mu} \mu_{\nu]} = 0 \quad , \quad 
    \p^{\mu} \nabla_{[\mu} \chi_{\nu]} = 0 \ , 
\eeq
where the square braces mean antisymmetrization of the enclosed indices. 
\ Comer \cite{C02} and Prix et al \cite{PCA02} discuss in some detail 
why the assumption of separate conservation for the two fluids should be 
reasonable for slow rotation and quasinormal mode calculations.

\subsection{The slow rotation expansions}

Andersson and Comer \cite{AC01c} have adapted to the superfluid case 
the ordinary fluid slow rotation scheme originally developed by Hartle 
\cite{H67}. \ The configurations are assumed to be axisymmetric, 
asymptotically flat, and stationary, with the metric taking the form  
\begin{eqnarray}
    g_{\mu \nu} {\rm d}x^{\mu} {\rm d}x^{\nu} &=& -\left(N^2 - 
      {\rm sin}^2\theta K \left[N^{\phi}\right]^2\right){\rm d}t^2 +  
      V {\rm d} r^2 - 2 {\rm sin}^2\theta K N^{\phi} {\rm d}t~ 
      {\rm d} \phi + \cr 
    && \cr
    &&K \left({\rm d}\theta^2 + {\rm sin}^2\theta {\rm d} \phi^2
      \right) \ . \label{finmetric}
\end{eqnarray}
The neutrons and protons are assumed to be rigidly rotating about the 
symmetry axis, with rates $\Omega_\n$ and $\Omega_\p$, respectively, 
and unit four-velocities written as 
\beq
    u_\n^{\mu} = \frac{t^{\mu} + \Omega_{\n} \phi^{\mu}}{\sqrt{N^2 - 
                 {\rm sin}^2 \theta K \left(N^{\phi} - \Omega_{\n}
                 \right)^2}} \qquad , \qquad
    u_\p^{\mu} = \frac{t^{\mu} + \Omega_{\p} \phi^{\mu}}{\sqrt{N^2 - 
                 {\rm sin}^2 \theta K \left(N^{\phi} - \Omega_{\p}
                 \right)^2}} \ ,
\eeq
where $t^{\mu}$ is the Killing vector associated with the stationarity, 
and $\phi^{\mu}$ with the axisymmetry.

The slow-rotation approximation assumes the rotation rates $\Omega_{\n}$ 
and $\Omega_{\p}$ are small in the sense that they should respect the 
inequalities (cf.~\cite{H67,AC01c})
\beq
    \Omega^2_{\n}~{\rm or}~\Omega^2_{\p}~{\rm or}~\Omega_{\n} 
    \Omega_{\p}~<<~\left(\frac{c}{R}\right)^2 \frac{G M}{R c^2} \ ,
\eeq  
where the speed of light $c$ and Newton's constant $G$ have been 
restored, and $M$ and $R$ are the mass and radius, respectively, of the 
non-rotating configuration. \ Because $G M/c^2 R < 1$, the inequalities 
naturally imply $\Omega_{\n} R << c$ and $\Omega_{\p} R << c$. \ The 
slow-rotation scheme has been shown, for instance in Prix et al 
\cite{PCA02}, to be a very good approximation for the fastest known 
pulsar, and starts to fail by about $15\%$ to $20\%$ for stars  
rotating at their Kepler limit. 

In the same manner as Hartle, the metric is expanded like 
\begin{eqnarray}
    N &=& e^{\nu(r)/2} \left(1 + h(r,\theta)\right) 
          \ , \cr
       && \cr 
    V &=& e^{\lambda(r)} \left(1 + 2 v(r,\theta)\right) 
          \ , \cr 
       && \cr
    K &=& r^2 (1 + 2 k(r,\theta)) \ , \cr
       && \cr
    N^{\phi} &=& \omega(r) \ , 
\end{eqnarray}
where $\omega$ is understood to be of ${\cal O}(\Omega_{\n,\p})$ and 
$h$, $v$, and $k$ of ${\cal O}(\Omega^2_{\n,\p})$. \ For later 
convenience we will also introduce
\beq
    \tilde{L}_{\n} = \omega - \Omega_{\n} \quad {\rm and} \quad 
    \tilde{L}_{\p} = \omega - \Omega_{\p} \ .
\eeq
The expansion for the neutron and proton number densities $\n$ and $\p$, 
respectively, are written as
\beq
    \n = \n_{\rm o}(r) \left(1 + \eta(r,\theta)\right) \qquad , \qquad 
    \p = \p_{\rm o} (r) \left(1 + \Phi(r,\theta) 
         \right) \ ,
\eeq
where the terms $\eta$ and $\Phi$ are understood to be of 
${\cal O}(\Omega^2_{\n,\p})$ and we have introduced the convention that 
terms with an ``${\rm o}$'' subscript are either contributions from the 
non-rotating background or quantities that are evaluated on the 
non-rotating background (e.g. $x^2_{\rm o} = \n_{\rm o} \p_{\rm o}$ 
etc.). \ One can show furthermore that the metric corrections $h$, $v$, 
and $k$ can be decomposed into  ``$l = 0$'' and ``$l = 2$'' terms using 
Legendre polynomials for the angular dependence, i.e. 
\begin{eqnarray}
    h &=& h_0(r) + h_2(r) P_2({\cos}\theta) \ , \cr
       && \cr
    v &=& v_0(r) + v_2(r) P_2({\cos}\theta) \ , \cr
       && \cr
    k &=& k_2(r) P_2({\cos}\theta) \ ,
\end{eqnarray}
where $P_2({\cos}\theta) = (3 {\cos}^2\theta - 1)/2$. \ As well we can 
write for the matter corrections  
\beq
    \eta = \eta_0(r) + \eta_2(r) P_2({\cos}\theta) 
             \quad , \quad
    \Phi = \Phi_0(r) + \Phi_2(r) P_2({\cos}\theta)  
             \ .
\eeq

Lastly, Andersson and Comer \cite{AC01c} have introduced a coordinate 
transformation $r \to r + \xi(r,\theta)$ that maps constant energy 
surfaces (i.e.~the level surfaces of $\Lambda_{\rm o}$) for the 
non-rotating background into the rotationally modified constant energy 
surfaces. \ The mapping $\xi$ is given as    
\beq
    \mu_{\rm o} \n_{\rm o} \eta_0 + \chi_{\rm o} \p_{\rm o} \Phi_0 + 
    \frac{r^2}{3 e^{\nu}} \A_{\rm o} \n_{\rm o} \p_{\rm o} 
    \left(\Omega_{\n} - \Omega_{\p}\right)^2 = 
    \Lambda^{\prime}_{\rm o} \xi_0 \label{coortrans0}
\eeq 
for $l = 0$ and for $l = 2$
\beq
    \mu_{\rm o} \n_{\rm o} \eta_2 + \chi_{\rm o} \p_{\rm o} \Phi_2 - 
    \frac{r^2}{3 e^{\nu}} \A_{\rm o} \n_{\rm o} \p_{\rm o} 
    \left(\Omega_{\n} - \Omega_{\p}\right)^2  = 
    \Lambda^{\prime}_{\rm o} \xi_2 \label{coortrans2}
\eeq
where $\mu_{\rm o} = \B_{\rm o} \n_{\rm o} + \A_{\rm o} \p_{\rm o}$ and 
$\chi_{\rm o} = \C_{\rm o} \p_{\rm o} + \A_{\rm o} \p_{\rm o}$.

\subsection{The slow-rotation equations}

No different in number, but different in form from the ordinary fluid 
case considered by Hartle, the Einstein/superfluid field equations 
reduce to four sets: (i) the non-rotating background that determines 
$\nu$, $\lambda$, $\n_{\rm o}$, and $\p_{\rm o}$, (ii) the linear order 
that calculates the ``frame-dragging'' $\omega$, (iii) the $l = 0$ 
second order equations that yield $\xi_0$, $\eta_0$, $\Phi_0$, $h_0$, 
and $v_0$, and (iv) the $l = 2$ set at second order that determines 
$\xi_2$, $\eta_2$, $\Phi_2$, $h_2$, $v_2$, and $k_2$. \ For convenience 
we list the equations here and recall where appropriate the free 
parameters of the system and how they are specified during a numerical 
integration.

\subsubsection{The ${\cal O}(\Omega_{\n,\p}^0)$ or background equations}

The background equations are
\begin{eqnarray}
  \lambda^{\prime} &=& \frac{1 - e^{\lambda}}{r} - 8 \pi r e^{\lambda} 
                       \Lambda_{\rm o} \ , \\
       && \cr
    \nu^{\prime} &=& - \frac{1 - e^{\lambda}}{r} + 8 \pi r e^{\lambda} 
                     \Psi_{\rm o} \ , \\
       && \cr
    0 &=& \left.\A^0_0\right|_{\rm o} \p_{\rm o}^{\prime} +
          \left.\B^0_0\right|_{\rm o} \n_{\rm o}^{\prime} + \frac{1}{2}
           \mu_{\rm o} \nu^{\prime} \ , \\ 
       && \cr
    0 &=& \left.\C^0_0\right|_{\rm o} \p_{\rm o}^{\prime} +
          \left.\A^0_0\right|_{\rm o} \n_{\rm o}^{\prime} + \frac{1}{2}
          \chi_{\rm o} \nu^{\prime} \ , \label{bckgrnd}
\end{eqnarray}
where the $\left.\a00\right|_{\rm o}$, $\left.\b00\right|_{\rm o}$, and 
$\left.\c00\right|_{\rm o}$ coefficients are obtained from the master 
function and will be discussed in much greater detail below 
(cf.~Secs.~\ref{anexp}, \ref{meanfield} and the appendix). \ Throughout 
a prime ``$~{}^{\prime}~$'' will denote differentiation with respect to 
$r$.

Regularity at the origin implies that $\lambda(0)$, 
$\lambda^{\prime}(0)$, $\nu^{\prime}(0)$, $\n_{\rm o}^{\prime}(0)$ and 
$\p_{\rm o}^{\prime}(0)$ all vanish. \ The radius $R$ is obtained from 
the condition that the pressure vanish on the surface 
(i.e.~$\Psi_{\rm o}(R) = 0$) and the background mass is given by 
\beq
    M = - 4 \pi \int_0^R \Lambda_{\rm o}(r) r^2 {\rm d} r \ . 
\eeq
The parameters $\n_{\rm o}(0)$ and $\p_{\rm o}(0)$ for the matter on the 
background are not independent because chemical equilibrium is imposed 
between the neutrons, protons, and a highly degenerate gas of 
relativistic electrons (cf.~the appendix).

\subsubsection{The ${\cal O}(\Omega_{\n,\p})$ or frame-dragging equation}

For ${\cal O}(\Omega_{\n,\p})$ there is only the frame-dragging 
$\omega(r)$, which is determined from
\beq
    \frac{1}{r^4} \left(r^4 e^{- (\lambda + \nu)/2} 
    \tilde{L}_{\n}^{\prime}\right)^{\prime} - 16 \pi e^{(\lambda - 
    \nu)/2} \left(\Psi_{\rm o} - \Lambda_{\rm o}\right) \tilde{L}_{\n} 
    = 16 \pi e^{(\lambda - \nu)/2} \chi_{\rm o} \p_{\rm o} 
    \left(\Omega_{\n} - \Omega_{\p}\right) \ . \label{frmdrg}
\eeq
The key difference with the frame-dragging equation for an ordinary 
fluid is the ``source'' term on the right-hand-side. \ Any solution must 
be such that it joins smoothly to the exterior vacuum solution. \ 
Continuity implies that 
\beq
    \tilde{L}_\n(R) = - \Omega_\n + \frac{2 J}{R^3} \ , 
\eeq
where $J$ is the total angular momentum of the system. \ Using also that 
the derivative at the surface is to be continuous implies  
\beq
    \tilde{L}_\n(R) = - \Omega_\n - \frac{R}{3} \tilde{L}^{\prime}_\n(R) 
                      \ . \label{con}
\eeq
When the frame-dragging equation is solved numerically, one searches for 
a central value $\tilde{L}_\n(0)$ that allows Eq.~(\ref{con}) to be 
satisfied. 

With a solution in hand, the neutron and proton angular momenta, $J_\n$ 
and $J_\p$, respectively, can then be calculated using
\beq
J_{\n} = - \frac{8 \pi}{3} \int_0^R {\rm d}r r^4 e^{(\lambda - \nu)/2}
         \left[\mu_{\rm o} \n_{\rm o} \tilde{L}_{\n} + \A_{\rm o}
         n_{\rm o} p_{\rm o} \left(\Omega_{\n} - \Omega_{\p}\right)
         \right]
\eeq
and 
\beq
J_{\p} = - \frac{8 \pi}{3} \int_0^R {\rm d}r r^4 e^{(\lambda - \nu)/2}
         \left[\chi_{\rm o} p_{\rm o} \tilde{L}_{\p} + \A_{\rm o}
         n_{\rm o} p_{\rm o} \left(\Omega_{\p} - \Omega_{\n}\right)
         \right] \ .
\eeq
The total angular momentum $J$ likewise follows since $J = J_\n + J_\p$.

\subsubsection{The ${\cal O}(\Omega^2_{\n,\p})$ equations}

One should first note that at this order there will exist both a $l = 0$ 
and $l = 2$ set of equations. \ The ${\cal O}(\Omega^2_{\n,\p})$, 
$l = 0$ set consists of Eq.~(\ref{coortrans0}) as well as
\begin{eqnarray}
    \gamma_{\n} &=& \frac{\left.\b00\right|_{\rm o} \n_{\rm o}}
                    {\mu_{\rm o}} \eta_0 + \frac{
                    \left.\a00\right|_{\rm o} \p_{\rm o}}{\mu_{\rm o}} 
                    \Phi_0 + \frac{r^2}{3 e^{\nu}} \frac{\p_{\rm o}}
                    {\mu_{\rm o}} \left(\A_{\rm o} + \n_{\rm o} 
                    \left.\frac{\partial \A}{\partial \n}\right|_{\rm o} 
                    + \n_{\rm o} \p_{\rm o} \left.\frac{\partial \A}
                    {\partial x^2}\right|_{\rm o}\right) 
                    \left(\Omega_{\n} - \Omega_{\p}\right)^2 - \cr
                 && \cr
                 && \frac{r^2}{3 e^{\nu}} \tilde{L}_{\n}^2 + h_0 \ , \\
                 && \cr
    \gamma_{\p} &=& \frac{\left.\c00\right|_{\rm o} \p_{\rm o}}
                    {\chi_{\rm o}} \Phi_0 + \frac{
                    \left.\a00\right|_{\rm o} \n_{\rm o}}{\chi_{\rm o}} 
                    \eta_0 + \frac{r^2}{3 e^{\nu}} \frac{\n_{\rm o}}
                    {\chi_{\rm o}} \left(\A_{\rm o} + \p_{\rm o} 
                    \left.\frac{\partial \A}{\partial \p}\right|_{\rm o} 
                    + \n_{\rm o} \p_{\rm o} \left.\frac{\partial \A}
                    {\partial x^2}\right|_{\rm o}\right) 
                    \left(\Omega_{\n} - \Omega_{\p}\right)^2 - \cr
                 && \cr
                 && \frac{r^2}{3 e^{\nu}} \tilde{L}_{\p}^2 + h_0 \ , \\
                    \label{flden0}
                 && \cr
   0 &=& \frac{16 \pi r^2}{3 e^{\nu}} \left[\left(\Psi_{\rm o} - 
         \Lambda_{\rm o}\right) \tilde{L}_{\n}^2 + \chi_{\rm o} 
         \p_{\rm o} \left(\Omega_{\n} - \Omega_{\p}\right) 
         \left(\tilde{L}_{\n} + \tilde{L}_{\p}\right) - \A_{\rm o} 
         \n_{\rm o} \p_{\rm o} \left(\Omega_{\n} - \Omega_{\p}
         \right)^2\right] + \cr
      && \cr
      && 8 \pi \Lambda^{\prime}_{\rm o} \xi_0 - \frac{2}{r^2} \left(
         \frac{r}{e^{\lambda}} v_0\right)^{\prime} + \frac{r^2}{6 e^{\nu 
         + \lambda}} \left(\tilde{L}_{\n}^{\prime}\right)^2 \ , \\
         \label{000}
      && \cr
   0 &=& \frac{2}{r e^{\lambda}} h^{\prime}_0 - \frac{2}{r e^{\lambda}} 
      \left(\nu^{\prime} + \frac{1}{r}\right) v_0 + \frac{r^2}{6 
      e^{\nu +\lambda}} \left(\tilde{L}_{\n}^{\prime}\right)^2 - 8 \pi 
      \left[\mu_{\rm o} \n_{\rm o} \gamma_{\n} + \chi_{\rm o} 
      \p_{\rm o} \gamma_{\p} - \left(\Psi_{\rm o} - \Lambda_{\rm o}
      \right) h_0 + \right. \cr
   && \cr
   && \left.\frac{r^2}{3 e^{\nu}} \left(\mu_{\rm o} \n_{\rm o} 
      \tilde{L}_{\n}^2 + \chi_{\rm o} \p_{\rm o} \tilde{L}_{\p}^2
      \right) - \frac{r^2}{3 e^{\nu}} \n_{\rm o} \p_{\rm o} \A_{\rm o} 
      \left(\Omega_{\n} - \Omega_{\p}\right)^2\right] \ . \label{110}
\end{eqnarray}
Here $\gamma_\n$ and $\gamma_\p$ are integration constants, which are 
determined from data given at the center of the star:
\begin{eqnarray}
    \gamma_{\n} &=& h_0(0) + \left.\left(\frac{\mu_{\rm o} \left.\a00
                  \right|_{\rm o} - \chi_{\rm o} \left.\b00
                  \right|_{\rm o}}{\mu_{\rm o}^2} \p_{\rm o} \Phi_0
                  \right)\right|_{r = 0} \ , \\
                 && \cr
    \gamma_{\p} &=& h_0(0) + \left.\left(\frac{\mu_{\rm o} \left.\c00
                  \right|_{\rm o} - \chi_{\rm o} \left.\a00
                  \right|_{\rm o}}{\mu_{\rm o} \chi_{\rm o}}
                  \p_{\rm o} \Phi_0\right)\right|_{r = 0} \ .
\end{eqnarray}
It is convenient to define the new variable $m_0 = r v_0/\exp(\lambda)$. 
\ Because we must have $\xi_0(0)=0$, then $\eta_0(0)$ and $\Phi_0(0)$ 
are related via
\beq
    \left.\left(\mu_{\rm o} \n_{\rm o} \eta_0\right)\right|_{r = 0} +
    \left.\left(\chi_{\rm o} \p_{\rm o} \Phi_0\right)\right|_{r = 0}
    = 0 \ .
\eeq
Regularity at the center of the star also requires that $m_0(0) = 0$, 
but $h_0(0)$ and $\eta_0(0)$, say, are freely specified.

As for the ${\cal O}(\Omega^2_{\n,\p})$, $l = 2$ equations, in addition 
to Eq.~(\ref{coortrans2}), we have 
\begin{eqnarray}
    0 &=& \frac{\left.\b00\right|_{\rm o} \n_{\rm o}}{\mu_{\rm o}} 
          \eta_2 + \frac{\left.\a00\right|_{\rm o} \p_{\rm o}}
          {\mu_{\rm o}} \Phi_2 - \frac{r^2}{3 e^{\nu}} \frac{\p_{\rm o}}
          {\mu_{\rm o}} \left(\A_{\rm o} + \n_{\rm o} 
          \left.\frac{\partial \A}{\partial \n}\right|_{\rm o} + 
          \n_{\rm o} \p_{\rm o} \left.\frac{\partial \A}{\partial x^2}
          \right|_{\rm o}\right) \left(\Omega_{\n} - \Omega_{\p}
          \right)^2 + \cr
       && \cr
       && \frac{r^2}{3 e^{\nu}} \tilde{L}_{\n}^2 + h_2 \ , \\ 
       && \cr
    0 &=& \frac{\left.\c00\right|_{\rm o} \p_{\rm o}}{\chi_{\rm o}} 
          \Phi_2 + \frac{\left.\a00\right|_{\rm o} \n_{\rm o}}
          {\chi_{\rm o}} \eta_2 - \frac{r^2}{3 e^{\nu}} 
          \frac{\n_{\rm o}}{\chi_{\rm o}} \left(\A_{\rm o} + \p_{\rm o} 
          \left.\frac{\partial \A}{\partial \p}\right|_{\rm o} + 
          \n_{\rm o} \p_{\rm o} \left.\frac{\partial \A}{\partial x^2}
          \right|_{\rm o}\right) \left(\Omega_{\n} - \Omega_{\p}
          \right)^2 + \cr
       && \cr
       && \frac{r^2}{3 e^{\nu}} \tilde{L}_{\p}^2 + h_2 \ , 
          \label{flden2} \\
       && \cr
    v_2 + h_2 &=& \frac{r^4}{6 e^{\nu + \lambda}} \left(
                  \tilde{L}_{\n}^{\prime}\right)^2 + \frac{8 \pi r^4}{3 
                  e^{\nu}} \left(\Psi_{\rm o} - \Lambda_{\rm o}\right) 
                  \tilde{L}_{\n}^2 + \frac{8 \pi r^4}{3 e^{\nu}} 
                  \left[\chi_{\rm o} \p_{\rm o} \left(\Omega_{\n} - 
                  \Omega_{\p}\right) \left(\tilde{L}_{\n} + 
                  \tilde{L}_{\p}\right) - \right.\cr
               && \cr
               && \left.\A_{\rm o} \n_{\rm o} \p_{\rm o} 
                  \left(\Omega_{\n} - \Omega_{\p}\right)^2\right] \ ,  
                  \label{rem} \\
               && \cr
     0 &=& \frac{1}{r} \left(v_2 + h_2\right) - \left(k_2 + h_2
           \right)^{\prime} - \frac{\nu^{\prime}}{2} \left(h_2 - v_2
           \right) \ , \label{122} \\
        && \cr
   0 &=& \frac{2}{r e^{\lambda}} h^{\prime}_2 - \frac{6}{r^2} h_2 - 
         \frac{2}{r e^{\lambda}} \left(\nu^{\prime} + \frac{1}{r}\right) 
         v_2 + \frac{1}{e^{\lambda}} \left(\nu^{\prime} + \frac{2}{r}
         \right) k^{\prime}_2 - \frac{4}{r^2} k_2 - \frac{r^2}{6 
         e^{\nu +\lambda}} \left(\tilde{L}_{\n}^{\prime}\right)^2 + \cr
   && \cr
   && 8 \pi \left[\left(\Psi_{\rm o} - \Lambda_{\rm o}\right) h_2 + 
      \frac{r^2}{3 e^{\nu}} \left(\mu_{\rm o} \n_{\rm o} 
      \tilde{L}_{\n}^2 + \chi_{\rm o} \p_{\rm o} \tilde{L}_{\p}^2
      \right) - \frac{r^2}{3 e^{\nu}} \n_{\rm o} \p_{\rm o} \A_{\rm o} 
      \left(\Omega_{\n} - \Omega_{\p}\right)^2\right]  \ . \label{112}
\end{eqnarray}
In contrast to the $l = 0$ case, the condition that $\xi_2(0) = 0 $ at 
the center of the star leads to $\eta_2(0) = 0 = \Phi_2(0)$. \ Also 
note that Eq.~(\ref{rem}) can be used to remove $v_2$ in terms of the 
functions $h_2$ and $\tilde{L}_{\n,\p}$. \ It is furthermore convenient 
to work with the new variable $\tilde{k} = h_2 + k_2$ \cite{H67}. \ 
Regularity at the center of the star implies that $h_2(r) \sim c_1 r^2$ 
and $\tilde{k}(r) \sim c_2 r^4$ as $r \to 0$, where the constants $c_1$ 
and $c_2$ are related by 
\beq
    c_1 + 2 \pi \left(\Psi_{\rm o}(0) - \frac{1}{3} \Lambda_{\rm o}(0)
    \right) c_2 = 0 \ .
\eeq

We can take advantage of an overall scale invariance of the field 
equations to remove the explicit dependence on both $\Omega_\n$ and 
$\Omega_\p$ by dividing through by $\Omega_\p$, the net result being 
that only the ratio $\Omega_\n/\Omega_\p$ needs to be specified in 
advance. \ As for the other parameters, they are chosen so that the 
solutions obtained join smoothly to a vacuum exterior. \ For $l = 0$,  
after $\Omega_\n/\Omega_\p$ and $\eta_0(0)$ are specified, this means a 
search over $h_0(0)$ is performed until a smooth solution is achieved. \ 
For $l = 2$, the situation is slightly more involved, requiring that  
homogeneous and then particular solutions be obtained for the set 
$(h_2,\tilde{k})$ (see \cite{AC01c} for more details). \ By searching 
over $c_1$, say, and also trying different linear combinations of 
homogeneous and particular solutions eventually smoothness with the 
vacuum exterior can be achieved.

Having obtained a complete solution, the rotationally induced change to 
the mass can be determined using  
\beq
    \delta M = \left(R - 2 M\right) v_0(R) + \frac{J^2}{R^3} \ .
\eeq 
Also the rotationally induced changes to the neutron and proton particle 
numbers are obtained, respectively, from
\beq
\delta N_n = \int_0^R {\rm d} r r^2 e^{\lambda/2} \n_{\rm o} 
             \left(\eta_0 + v_0 + \left[\frac{\lambda^{\prime}}{2} + 
             \frac{2}{r}\right] \xi_0 + \frac{2 r^2}{3 e^{\nu}} 
             \tilde{L}^2_{\n}\right) \label{baryn}
\eeq
and
\beq
 \delta N_p = \int_0^R {\rm d} r r^2 e^{\lambda/2} \p_{\rm o} 
              \left(\Phi_0 + v_0 + \left[\frac{\lambda^{\prime}}{2} + 
              \frac{2}{r}\right] \xi_0 + \frac{2 r^2}{3 e^{\nu}} 
              \tilde{L}^2_{\p}\right) \ . \label{baryp}
\eeq
Finally, the Kepler frequency $\Omega_K$ is calculated 
(cf.~\cite{FIP86} and \cite{AC01c}) using the slow-rotation form of 
\beq
    \Omega_K = \frac{N v}{\sqrt{K}} + \omega
\eeq
where the metric variables are those of Eq.~(\ref{finmetric}) and
\beq
    v = \frac{K^{3/2} \omega^\prime}{N K^\prime} + \sqrt{
        \frac{2 K N^\prime}{N K^\prime} + \left(\frac{K^{3/2} 
        \omega^\prime}{N K^\prime}\right)^2}
\eeq
is the orbital velocity according to a zero-angular momentum observer 
at the equator (where all quantities are to be evaluated). \ In the slow 
rotation approximation, the Kepler limit is the solution to 
\beq
     \Omega_K = \sqrt{\frac{M}{R^3}} - \frac{\hat{J} \Omega_\p}{R^3} + 
                \sqrt{\frac{M}{R}} \left\{\frac{\delta \hat{M}}{2M} + 
                \frac{(R + 3 M) (3 R - 2 M)}{4 R^4 M^2} \hat{J}^2 -
                \frac{3}{4} \frac{2 \hat{\xi}_0 + \hat{\xi}_2}{R^2} + 
                \alpha \hat{A} \right\} \Omega_\p^2 \label{kep3}
\eeq
where we have made the scaling with $\Omega_p$ explicit by introducing
$J = \hat{J} \Omega_p$, etc., and have also defined
\beq
    \alpha = \frac{3 (R^3 - 2 M^3)}{4 M^3} \log \left(1 - \frac{2 M}{R} 
             \right) + \frac{3 R^4 - 3 R^3 M - 2 R^2 M^2 - 8 R M^3 + 6 
             M^4}{2 R M^2(R - 2 M)} \ .
\eeq
In solving Eq.~\ref{kep3} we fix first the ratio $\Omega_\n/\Omega_\p$. 
\ If $|\Omega_\n/\Omega_\p| < 1$, then we insert $\Omega_K = \Omega_p$ 
and solve for $\Omega_\p$, otherwise we set $\Omega_K = 
(\Omega_\n/\Omega_\p) \Omega_\p$ and then again solve for $\Omega_\p$.

\subsection{Analytic expansion for the local matter content} 
\label{anexp}

Before proceeding to the next section that describes the mean field 
approach, one final observation needs to be made. \ We see in the 
equations above the appearance of not only the entrainment, via the $\A$ 
coefficient, but also the relative velocity-dependent part of the 
entrainment, through the term $\partial \A / \partial x^2$. \ A priori 
one needs a formalism for the matter that will determine this term, if 
the effects of a relative rotation between the neutrons and protons are 
to be consistently incorporated. \ The mean field formalism presented 
in Sec.~\ref{meanfield} can do just that. \ However, before embarking on 
a full-fledged calculation, it is very worthwhile to consider next the 
implications of the slow rotation expansion for the master function. 

The term $\partial \A / \partial x^2$ would seem to imply, at least in 
principle, that we need to know the master function to ${\cal O}(x^4)$. 
\ Consider, then, regions of the matter (i.e.~fluid elements) that are 
local enough that their spacetime can be taken to be that of Minkowski, 
so that we can write out the $x^2$ term explicitly as  
\beq
    x^2 = \n \p \left(\frac{1 - \vec{v}_{\n} \cdot \vec{v}_{\p}/c^2} 
          {\sqrt{1 - (v_{\n}/c)^2} \sqrt{1 - (v_{\p}/c)^2}}\right) \ ,
\eeq
where $\vec{v}_\n$ and $\vec{v}_\p$ are the neutron and proton, 
respectively, three-velocities in the local Minkowski frame. \ We see 
from this that when the individual three-velocities $\vec{v}_{\n ,\p}$ 
satisfy $v_{\n,\p} / c << 1$, then $x^2 \approx \n \p$ to leading order 
in the ratios $v_{\n}/c$ and $v_{\p}/c$. 

With this as motivation we write an analytic expansion for the master 
function of the form 
\cite{ACL02,CJ03}
\beq
    \Lambda(n^2,p^2,x^2) = \sum_{i = 0}^{\infty} \lambda_i(n^2,p^2) 
                           \left(x^2 - \n \p\right)^i \ .
\eeq
The $\A$, $\a00$, etc.~coefficients that appear in the field equations 
can now be written as 
\begin{eqnarray}
  \A &=& - \lambda_1 - \sum_{i = 2}^{\infty} i~\lambda_i \left(x^2 - 
         \n \p\right)^{i - 1}  \ , \\
     && \cr
  \B &=& - \frac{1}{\n} \frac{\partial \lambda_0}{\partial \n} - 
         \frac{\p}{\n} \A - \frac{1}{\n} \sum_{i = 1}^{\infty} 
         \frac{\partial \lambda_i}{\partial \n} \left(x^2 - \n \p
         \right)^i \ , \\
     && \cr
  \C &=& - \frac{1}{\p} \frac{\partial \lambda_0}{\partial \p} - 
         \frac{\n}{\p} \A - \frac{1}{\p} \sum_{i = 1}^{\infty} 
         \frac{\partial \lambda_i}{\partial \p} \left(x^2 - \n \p
         \right)^i \ , \\
     && \cr
  \a00 &=& - \frac{\partial^2 \lambda_0}{\partial \p \partial \n} - 
           \sum_{i = 1}^{\infty} \frac{\partial^2 \lambda_i}{\partial 
           \p \partial \n} \left(x^2 - \n \p\right)^i \ , \\
     && \cr
  \b00 &=& - \frac{\partial^2 \lambda_0}{\partial \n^2} - 
           \sum_{i = 1}^{\infty} \frac{\partial^2 \lambda_i}{\partial 
           \n^2} \left(x^2 - \n \p\right)^i \ , \\
     && \cr
  \c00 &=& - \frac{\partial^2 \lambda_0}{\partial \p^2} - 
           \sum_{i = 1}^{\infty} \frac{\partial^2 \lambda_i}{\partial 
           \p^2} \left(x^2 - \n \p\right)^i \ , \\
        && \cr
  \frac{\partial \A}{\partial \n} &=& - \frac{\partial \lambda_1}
           {\partial \n} - \sum_{i = 2}^{\infty} i \left(\frac{\partial 
           \lambda_i}{\partial \n} \left[x^2 - \n \p\right] - 
           \left[i - 1\right] \p \lambda_i\right) \left(x^2 - \n \p
           \right)^{i - 2} \ , \\   
        && \cr
  \frac{\partial \A}{\partial \p} &=& - \frac{\partial \lambda_1} 
           {\partial \p} - \sum_{i = 2}^{\infty} i \left(\frac{\partial 
           \lambda_i}{\partial \p} \left[x^2 - \n \p\right] - 
           \left[i - 1\right] \n \lambda_i\right) \left(x^2 - \n \p
           \right)^{i - 2} \ , \\   
  \frac{\partial \A}{\partial x^2} &=& - 2 \lambda_2 - 
           \sum_{i = 3}^{\infty} i \left(i - 1\right) \lambda_i 
           \left(x^2 - \n \p\right)^{i - 2} \ .
\end{eqnarray}

We conclude from this expansion that the master function coefficients 
$\lambda_0$, $\lambda_1$, and $\lambda_2$ uniquely determine the 
background values for all the matter coefficients---i.e.~$\A_{\rm o}$, 
$\B_{\rm o}$, etc---that enter the field equations. \ In particular, if 
we need to know $\lambda_2$ then we need to know the master function to 
at least ${\cal O}(x^4)$. \ However, the particular combinations that 
contain $\partial \A / \partial x^2$ that enter the field equations are 
seen to reduce to
\begin{eqnarray}
    \A + \n \frac{\partial \A}{\partial \n} + \n \p \frac{\partial \A}
         {\partial x^2} &=& - \lambda_1 - \n \frac{\partial \lambda_1}
         {\partial \n} - \sum_{i = 2}^\infty \left(\lambda_i + \n 
         \frac{\partial \lambda_i}{\partial \n}\right) \left(x^2 - \n 
         \p\right)^{i - 1} \ , \\
    && \cr
    \A + \p \frac{\partial \A}{\partial \p} + \n \p \frac{\partial \A}
         {\partial x^2} &=& - \lambda_1 - \p \frac{\partial \lambda_1}
         {\partial \p} - \sum_{i = 2}^\infty \left(\lambda_i + \p 
         \frac{\partial \lambda_i}{\partial \p}\right) \left(x^2 - \n 
         \p\right)^{i - 1} \ .
\end{eqnarray} 
That is, these combinations involve only the $\lambda_1$ coefficient 
when evaluated on the background. \ This, in fact, makes perfect sense 
since the slow rotation approximation is to 
${\cal O}(\Omega^2_{\n,\p})$, which naturally corresponds to the term 
in the master function proportional to $x^2 - \n \p$.

\section{The $\sigma$ - $\omega$ Relativistic Mean Field Model} 
\label{meanfield}

The master function to be used has been obtained using a relativistic 
$\sigma - \omega$ mean field model of the type described by 
Glendenning \cite{G97}. \ The Lagrangian for this system is given by 
\beq
    L = L_{b} + L_{\sigma} + L_{\omega} + L_{int} \ ,
\eeq
where
\begin{eqnarray}
   L_{b} &=& \bar{\psi} (i\gamma _{\mu } \partial^{\mu } - m) \psi \ , 
             \\
          && \cr
   L_{\sigma} &=& - \frac{1}{2} \partial _{\mu} \sigma \partial^{\mu} 
                   \sigma -\frac{1}{2}m_{\sigma }^{2} \sigma ^{2} \ , \\
               && \cr
   L_{\omega} &=& - \frac{1}{4} \omega _{\mu \nu} \omega ^{\mu \nu}
                  - \frac{1}{2} m_{\omega}^{2} \omega _{\mu} 
                  \omega ^{\mu} \ , \\
               && \cr
   L_{int} &=& g_{\sigma} \sigma \bar{\psi} \psi - g_{\omega} 
               \omega _{\mu } \bar{\psi} \gamma^{\mu }\psi \ ,
\end{eqnarray}
Here $m$ is the baryon mass, $\psi$ is an 8-component spinor with the 
proton components as the top 4 and the neutron components as the bottom 
4, the $\gamma _{\mu}$ are the corresponding $8 \times 8$ block diagonal 
Dirac matrices, and $\omega _{\mu \nu} = \partial _{\mu} \omega _{\nu} - 
\partial_{\nu} \omega _{\mu }$. \ The coupled set of field equations 
obtained from this Lagrangian are to be solved in each fluid element of 
the neutron star. \ The main approximations of the mean field approach 
are to assume that the nucleons can be represented as plane-wave states 
and that all gradients of the $\sigma$ and $\omega^\mu$ fields can be 
ignored. \ The coupling constants $g_\sigma$ and $g_\omega$ and field 
masses $m_\sigma$ and $m_\omega$ are determined, for instance, from 
properties of nuclear matter at the nuclear saturation density. \ 
Fortunately, in what follows, we only need to give the ratios 
$c^2_\sigma = (g_\sigma/m_\sigma)^2$ and $c^2_\omega = 
(g_\omega/m_\omega)^2$.

\begin{figure}[t]
\centering
\vskip 24pt
\includegraphics[height=7.5cm,clip]{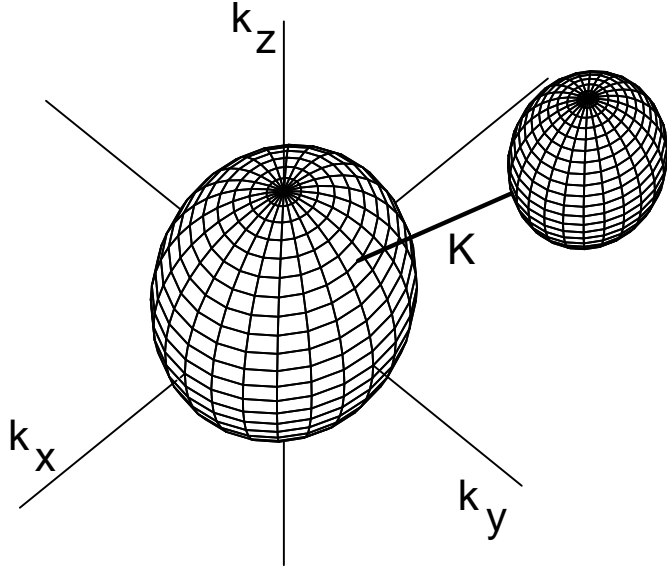}
\caption{The neutron and proton Fermi spheres drawn in momentum space. 
\ The neutron sphere is centered at the origin and has radius $k_\n$, 
whereas the proton sphere has radius $k_\p$ and its center has been 
displaced an amount $K$ from the origin.}
\label{cartoon}
\end{figure}

Of course, the main consideration is to produce a master function that 
incorporates the entrainment effect. \ Consider again fluid elements  
somewhere in the neutron star. \ The fermionic nature of the nucleons 
means that they are to be placed into the various energy levels 
(obtained from the mean field calculation) until their respective 
(local) Fermi spheres are filled. \ The Fermi spheres are best 
visualized in momentum space, as in Fig.~\ref{cartoon}. \ The 
entrainment is incorporated by displacing the center of the proton 
Fermi sphere from that of the neutron Fermi sphere, also illustrated in 
Fig.~\ref{cartoon}. \ The neutron sphere is centered on the origin, and 
has a radius $k_\n$. \ Displaced an amount $K$ from the origin is the 
center of the proton sphere, which has a radius $k_\p$. \ The Fermi 
sphere radii and displacement $(k_\n,k_\p,K)$ are functions of the local 
neutron and proton number densities and the local, relative velocity of 
the protons, say, with respect to the neutrons. 

Introducing the definitions  
\beq
    \phi_\n \equiv g_\omega \omega^z \quad , \quad 
    \phi_\p \equiv \phi_\n + K \ ,
\eeq
then the master function takes the form \cite{CJ03}
\beq
    \Lambda = - \frac{c^2_{\omega}}{18 \pi^4} \left(k^3_\n + k^3_\p
    \right)^2 + \frac{1}{2 c^2_{\omega}} \phi^2_\n - \frac{1}{2 
    c^2_{\sigma}} \left(m^2 - m_*^2\right) - 3 \langle \bar{\Psi} 
    \gamma^x k_x \Psi \rangle \ ,
\eeq
where
\begin{eqnarray}
    \langle \bar{\Psi} \gamma^x k_x \Psi \rangle &=& \frac{1}{12 \pi^2} 
            \left(\int_{- k_\n}^{k_\n} d k_z \left[\left(k^2_\n
            - 2 m^2_* - 2 \phi^2_\n - 3 k^2_z - 4 \phi_\n k_z\right) 
            \left(k_\n^2 + \phi_\n^2 + m_*^2 + 2 \phi_\n k_z
            \right)^{1/2} \right.\right. \cr
         && \cr
         && \left.\left.+ 2 \left(\left[k_z + \phi_\n\right]^2 + m_*^2
            \right)^{3/2}\right] + \int_{- k_\p}^{k_\p} d k_z \left[
            \left(k^2_\p - 2 m^2_* - 2 \phi_\p^2 - 3 k^2_z - 4 
            \phi_\p k_z\right) \right.\right.\cr
         && \cr
         && \left.\left.\left(k_\p^2 + \phi_\p^2 + m_*^2 + 2 \phi_\p 
            k_z\right)^{1/2} + 2 \left(\left[k_z + \phi_\p\right]^2 + 
            m_*^2\right)^{3/2}\right]\right) \ , \\
         && \cr
    m_* &=& m - \frac{c^2_{\sigma}}{2 \pi^2} m_* \left(
            \int_{- k_\n}^{k_\n} d k_z \left[k_\n^2 + \phi_\n^2 + 
            m_*^2 + 2 \phi_\n k_z\right]^{1/2} + \right. \cr
         && \cr 
         && \left.\int_{- k_\p}^{k_\p} d k_z \left[k_\p^2 + \phi_\p^2 + 
            m_*^2 + 2 \phi_\p k_z\right]^{1/2} - \int_{- k_\n}^{k_\n} 
            d k_z \left[\left(k_z + \phi_\n\right)^2 + m_*^2\right]^{1/2}
            - \right. \cr
         && \cr
         && \left.\int_{- k_\p}^{k_\p} d k_z \left[\left(k_z + \phi_\p
            \right)^2 + m_*^2\right]^{1/2}\right) \ , \label{mstar} \\
         && \cr
    \n^{0} &=& \frac{1}{3 \pi^2} k_{\n}^3 \ , \label{n0} \\
         && \cr
    \p^{0} &=& \frac{1}{3 \pi^2} k_{\p}^3 \ , \label{p0} \\
         && \cr
    \n^{z} &=& \frac{1}{2 \pi^2} \int_{- k_{\n}}^{k_{\n}} d k_{z} 
               \left(k_{z} + \phi_\n\right) \left(\left[k_{\n}^2 + m_*^2 
               + \phi_\n^2 + 2 \phi_\n k_{z}\right]^{1/2} - \right.\cr
            && \cr
            && \left.\left[\left(k_{z} + \phi_\n\right)^2 + m_*^2
               \right]^{1/2}\right) \ , \label{nz} \\
         && \cr
    \p^{z} &=& \frac{1}{2 \pi^2} \int_{- k_{\p}}^{k_{\p}} d k_{z} 
               \left(k_{z} + \phi_\p\right) \left(\left[k_{\p}^2 + m_*^2 
               + \phi_\p^2 + 2 \phi_\p k_{z}\right]^{1/2} - \right.\cr
            && \cr
            && \left.\left[\left(k_{z} + \phi_\p\right)^2 + m_*^2
               \right]^{1/2}\right) \ , \label{pz} \\ 
         && \cr
    \phi_\n &=& - c_{\omega}^2 \left(\n^z + \p^z\right) \ . 
               \label{phin}
\end{eqnarray}
As applied to this system, the slow-rotation approximation means that 
$K$ is small with respect to $k_{\n, \p}$. \ In particular, we should 
keep terms up to and including ${\cal O}(K^2)$.

Implementation of the mean-field approach is hampered by the fact that 
the most natural variables for it are the momenta $(k_\n,k_\p,K)$, 
whereas the general relativistic superfluid formalism relies on  
$(\n^2,\p^2,x^2)$. \ The two sets are related to each other via the 
kinematic relations
\beq
    \n^2 = \left(\n^0\right)^2 - \left(\n^z\right)^2 \quad , \quad
    \p^2 = \left(\p^0\right)^2 - \left(\p^z\right)^2 \quad , \quad
    x^2 = \n^0 \p^0 - \n^z \p^z \ , \label{npx}
\eeq
where $\n^0$, $\p^0$, $\n^z$, and $\p^z$ are given above in 
Eqs.~(\ref{n0}), (\ref{p0}), (\ref{nz}), and (\ref{pz}). \ In principle 
one would specify values for the set $(\n^2,\p^2,x^2)$, and then use 
Eqs.~(\ref{n0}), (\ref{p0}), (\ref{nz}), and (\ref{pz}) in 
Eq.~(\ref{npx}) to determine the set $(k_\n,k_p,K)$. \ One would then 
solve Eq.~(\ref{mstar}) for the Dirac effective mass $m_*$. \ In 
practice we rewrite the field equations so as to solve directly for 
$(k_\n,k_\p,K)$.

Earlier we discussed the slow-rotation approximation as it applies to 
the master function, and determined that we need to know the 
$\lambda_0$ and $\lambda_1$ coefficients. \ Comer and Joynt \cite{CJ03} 
have shown that
\begin{eqnarray}
   \lambda_0 &=& - \frac{c^2_{\omega}}{18 \pi^4} \left(k_\n^3 + k_\p^3
                 \right)^2 - \frac{1}{4 \pi^2} \left(k_\n^3 \sqrt{k_\n^2 
                 + \ms0^2} + k_\p^3 \sqrt{k_\p^2 + \ms0^2}\right) - \cr
              && \cr
              && \frac{1}{4 c^2_{\sigma}} \left(2 m - \ms0\right) \left(
                 m - \ms0\right) -  \frac{1}{8 \pi^2} \left(m_e k_\p 
                 \left[2 k_\p + m_e\right] \sqrt{k_\p^2 + m^2_e} - 
                 \right.\cr
              && \cr
              && \left.m^4_e {\rm ln}\left[\frac{k_\p + \sqrt{k^2_\p + 
                 m^2_e}}{m_e}\right]\right) \ , \\
              && \cr
   \lambda_1 &=& - c_{\omega}^2 - \frac{c^2_{\omega}}{5 \mu^2_{\rm o}} 
                 \left(2 k^2_\p \frac{\sqrt{k^2_\n + \ms0^2}} 
                 {\sqrt{k^2_\p + \ms0^2}} + \frac{c^2_{\omega}}{3 \pi^2} 
                 \left[\frac{k^2_\n k^3_\p}{\sqrt{k^2_\n + \ms0^2}} + 
                 \frac{k^2_\p k^3_\n}{\sqrt{k^2_\p + \ms0^2}}\right]
                 \right) - \cr
              && \cr
              && \frac{3 \pi^2 k^2_{\p}}{5 \mu^2_{\rm o} k^3_\n} 
                 \frac{k^2_\n + \ms0^2}{\sqrt{k^2_\p + \ms0^2}} \ ,
\end{eqnarray}
where $\ms0$ is the solution to the transcendental equation 
\begin{eqnarray}
  \ms0 &=& m - \ms0 \frac{c^2_{\sigma}}{2 \pi^2} \left(k_\n \sqrt{k_\n^2
            + \ms0^2} + k_\p \sqrt{k_\p^2 + \ms0^2} - \right.\cr
        && \cr
        && \left.\frac{1}{2} \ms0^2 {\rm ln} \left[\frac{k_\n + 
           \sqrt{k_\n^2  + \ms0^2}}{- k_\n + \sqrt{k_\n^2 + \ms0^2}}
           \right] - \frac{1}{2} \ms0^2 {\rm ln} \left[\frac{k_\p + 
           \sqrt{k_\p^2 + \ms0^2}}{- k_\p + \sqrt{k_\p0^2 + \ms0^2}}
           \right]\right) \label{diracmass}
\end{eqnarray}
and we have added to $\lambda_0$ the contributions due to the electrons 
($m_e = m / 1836$). \ This is necessary to obtain a central proton 
fraction $\p_{\rm o}(0)/(\p_{\rm o}(0) + \n_{\rm o}(0))$ of about 
$0.1$, which is considered typical for neutron stars.

In order to solve for the background, we note that 
\beq
    \n_{\rm o} = \frac{k_\n^3}{3 \pi^2} \quad , \quad 
    \p_{\rm o} = \frac{k_\p^3}{3 \pi^2} \ .
\eeq
That is, we replace everywhere the background neutron and proton 
number densities with their respective Fermi momenta. \ Once the Fermi 
momenta are specified, we can then determine the background value for 
the Dirac effective mass, i.e.~$\ms0$, from Eq.~(\ref{diracmass}). \ 
However, Comer and Joynt found that numerical solutions were easier to 
obtain by turning Eq.~(\ref{diracmass}) into a differential equation 
using the identity
\beq
    \ms0^{\prime} = \frac{\partial \ms0}{\partial k_\n} k^{\prime}_\n + 
                    \frac{\partial \ms0}{\partial k_\p} k^{\prime}_\p \ ,
                    \label{msprime}
\eeq
and where $k^{\prime}_\n$ and $k^{\prime}_\p$ are to be obtained 
simultaneously using Eq.~(\ref{bckgrnd}). \ The partial derivatives of 
the Dirac effective mass can be found in the appendix, along with the 
other matter coefficients.

\section{Numerical Results} \label{results}

In this section we present numerical solutions to the slow rotation 
equations. \ The two canonical background configurations (see Table 
\ref{table1}) that will be used repeatedly are the same as those used 
by Comer and Joynt \cite{CJ03}. \ They are such that the neutron and 
proton number densities vanish on the same surface. \ They are also such 
that chemical equilibrium has been imposed, which implies that the 
proton number density at the center of the star is no longer a free 
parameter. \ We do not apply chemical equilibrium for the rotating 
configurations since Andersson and Comer \cite{AC01c} have shown it is 
not consistent with rigid rotation. \ The two sets of parameter values 
for the mean field model given in Table \ref{table1} represent the two 
extremes discussed by Glendenning \cite{G97}. \ For all of the solutions 
we will set $\eta_0(0) = 0$ which implies also that $\Phi_0(0) = 0$. 

\begin{table}
\begin{tabular}{|c|c|}
\hline
  Model I & Model II \cr
\hline
  $\left(g_\sigma/m_\sigma\right)^2 = 12.684$ & $\left(g_\sigma/m_\sigma
  \right)^2 =8.403$ \cr  $\left(g_\omega/m_\omega\right)^2 = 7.148$ & 
  $\left(g_\omega/m_\omega\right)^2 = 4.233$ \cr
  $\nu(0) = - 2.316408$ & $\nu(0) = - 2.288385$ \cr
  $k_\n(0) = 2.8~{\rm fm}^{- 1}$ & $k_\n(0) = 3.25~{\rm fm}^{- 1}$ \cr
  $x_\p(0) = 0.101$ & $x_\p(0) = 0.102$ \cr
  $M = 2.509~M_\odot$ & $M = 1.996~M_\odot$ \cr
  $R = 11.696~{\rm km}$ & $R = 9.432~{\rm km}$ \cr 
  $\eta_0(0) = 0$ & $\eta_0(0) = 0$ \cr
\hline
\end{tabular}
\caption{The canonical background models used here and by Comer and 
Joynt \cite{CJ03}.}
\label{table1}
\end{table}

\begin{figure}[t]
\centering
\vskip 24pt
\includegraphics[height=6.5cm,clip]{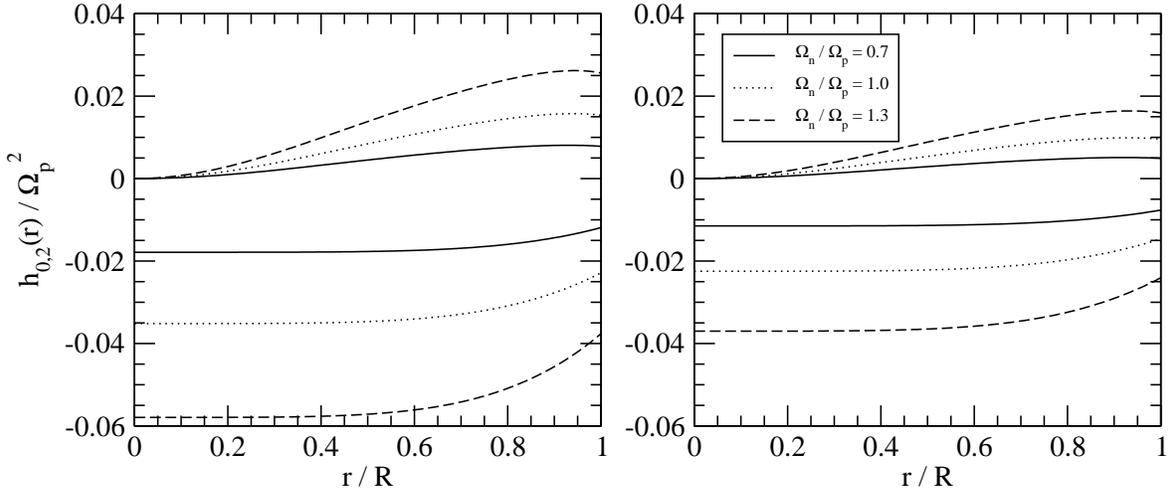}
\caption{The metric functions $h_0(r)$---the 3 lower curves---and 
$h_2(r)$---the 3 upper curves---vs $r$ for Models I (left panel) and II 
(right panel) of Table~\ref{table1}. \ The relative rotation rates are 
varied ($\Omega_\n/\Omega_\p = 0.7, 1.0, 1.3$).}
\label{h0}
\end{figure}

\begin{figure}[t]
\centering
\vskip 24pt
\includegraphics[height=6.5cm,clip]{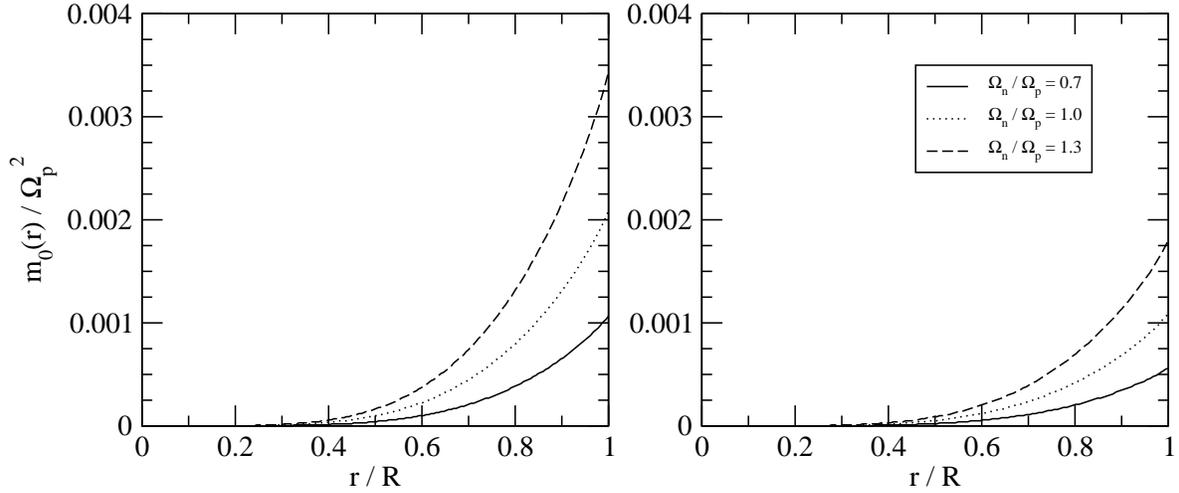}
\caption{The metric function $m_0(r) = r v_0(r)/\exp(\lambda(r))$ vs $r$ 
for Models I (left panel) and II (right panel) of Table~\ref{table1}. \ 
The relative rotation rates are varied ($\Omega_\n/\Omega_\p = 0.7, 1.0, 
1.3$).}
\label{m0}
\end{figure}

\begin{figure}[t]
\centering
\vskip 24pt
\includegraphics[height=6.5cm,clip]{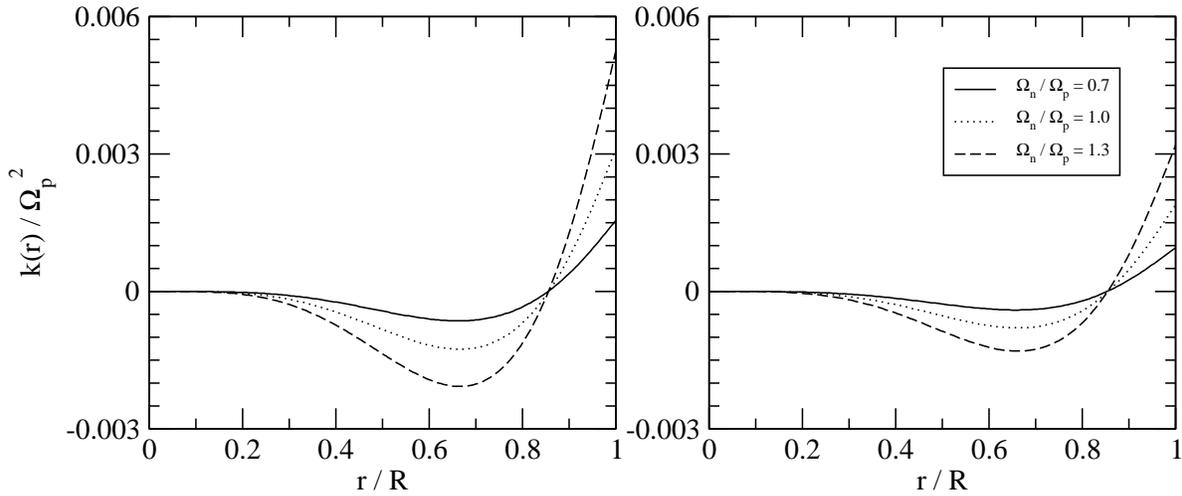}
\caption{The combination $\tilde{k}(r) = h_2(r) + k_2(r)$ vs $r$ for 
Models I (left panel) and II (right panel) of Table~\ref{table1}. \ The 
relative rotation rates are varied ($\Omega_\n/\Omega_\p = 0.7, 1.0, 
1.3$).}
\label{ktilde}
\end{figure}

\begin{figure}[t]
\centering
\vskip 24pt
\includegraphics[height=6.5cm,clip]{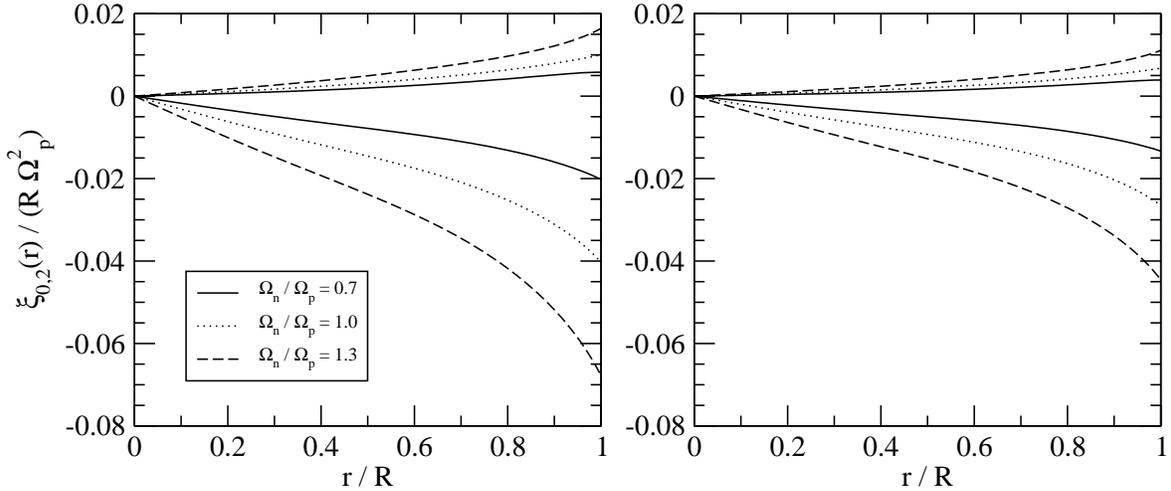}
\caption{The radial displacements $\xi_0(r)$---the 3 upper curves---and 
$\xi_2(r)$---the 3 lower curves---vs $r$ for Models I (left panel) and 
II (right panel) of Table~\ref{table1}. \ The relative rotation rates 
are varied ($\Omega_\n/\Omega_\p = 0.7, 1.0, 1.3$).}
\label{xi}
\end{figure}

\begin{figure}[t]
\centering
\vskip 24pt
\includegraphics[height=6.5cm,clip]{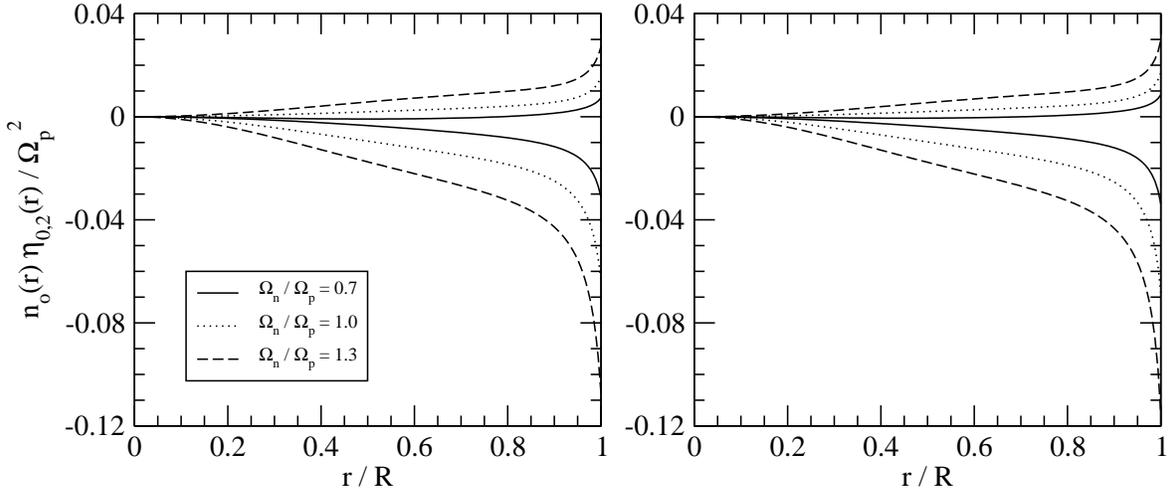}
\caption{The neutron density corrections $\n_{\rm o}(r) \eta_0(r)$---the 
3 upper curves---and $\n_{\rm o}(r) \eta_2(r)$---the 3 lower curves---vs 
$r$ for Models I (left panel) and II (right panel) of 
Table~\ref{table1}. \ The relative rotation rates are varied 
($\Omega_\n/\Omega_\p = 0.7, 1.0, 1.3$).}
\label{eta}
\end{figure}

\begin{figure}[t]
\centering
\vskip 24pt
\includegraphics[height=6.5cm,clip]{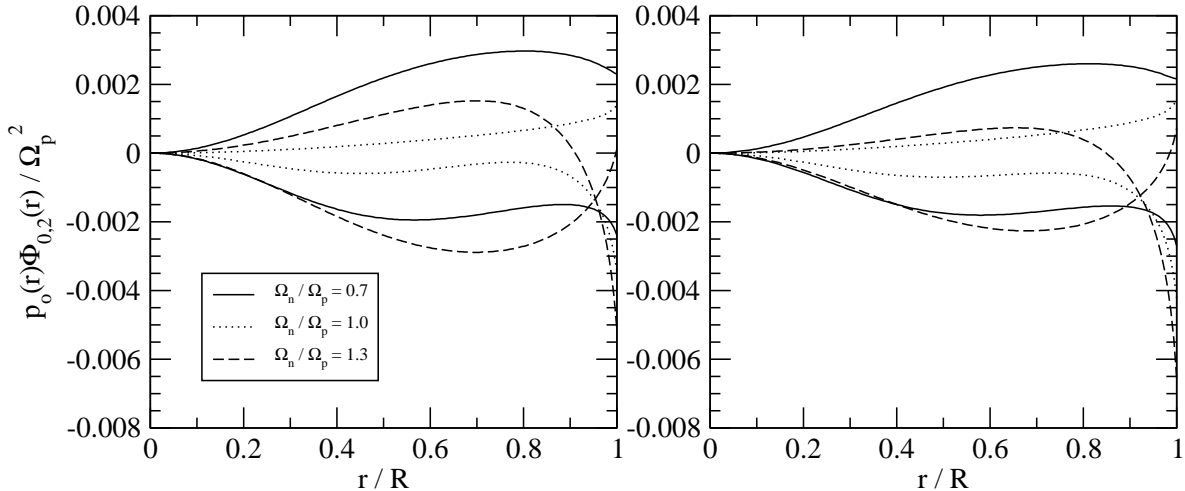}
\caption{The proton density corrections $\p_{\rm o}(r) \Phi_0(r)$---the 
3 uppermost curves at $r/R = 1$---and $\p_{\rm o}(r) \Phi_2(r)$---the 
3 lowermost curves at $r/R = 1$---vs $r$ for Models I (left panel) and 
II (right panel) of Table~\ref{table1}. \ The relative rotation rates 
are varied ($\Omega_\n/\Omega_\p = 0.7, 1.0, 1.3$).}
\label{phi}
\end{figure}

To begin we will give plots of the various field radial profiles for 
varying relative rotation rates of $\Omega_\n / \Omega_\p = 0.7,1.0,1.3$.
\ These fall within the range of rates that were considered earlier by 
Comer and Joynt \cite{CJ03}, who determined the frame-dragging and the 
angular momenta (cf.~their Figs.~4 and 5). \ Picking up at the second 
order, we present in Fig.~\ref{h0} a plot of $h_0(r)$ vs $r$ for several 
relative rotation rates and for Models I (left panel) and II (right 
panel) of Table \ref{table1}. \ Similarly in Fig.~\ref{m0} we show a 
plot of $m_0(r) = r v_0(r)/\exp(\lambda(r))$ vs $r$ and in 
Fig.~\ref{ktilde} we have $\tilde{k}(r) = h_2(r) + k_2(r)$ vs $r$. \ 
Fig.~\ref{xi} shows the radial profiles of $\xi_0(r)$ and $\xi_2(r)$. \ 
Figs.~\ref{eta} and \ref{phi} show the radial profiles of $\eta_0(r)$ 
and $\eta_2(r)$ and $\Phi_0(r)$ and $\Phi_2(r)$, respectively. \ (We do 
not show plots of $v_2(r)$ since it can be removed in terms of $h_2(r)$ 
and $\tilde{L}_{\n,\p}(r)$.) 

Andersson and Comer \cite{AC01c} used a simplified equation of state 
(i.e.~a sum of polytropes) and did not include entrainment and thus a 
direct comparison between our solutions and theirs would not be expected 
to yield the same details. \ That being said, there are some qualitative 
similarities. \ For instance, the signs of the $l = 0$ fields versus the 
$l = 2$ fields are in agreement, as are the number of zeroes that occur 
in the fields. \ The most pronounced qualitative differences are found 
in the functions $\Phi_0(r)$ and $\Phi_2(r)$, especially near the 
surface of the star. \ This could perhaps be related to our use of an 
explicit term for the electrons in the equation of state.  

\begin{figure}[t]
\centering
\vskip 24pt
\includegraphics[height=8.0cm,clip]{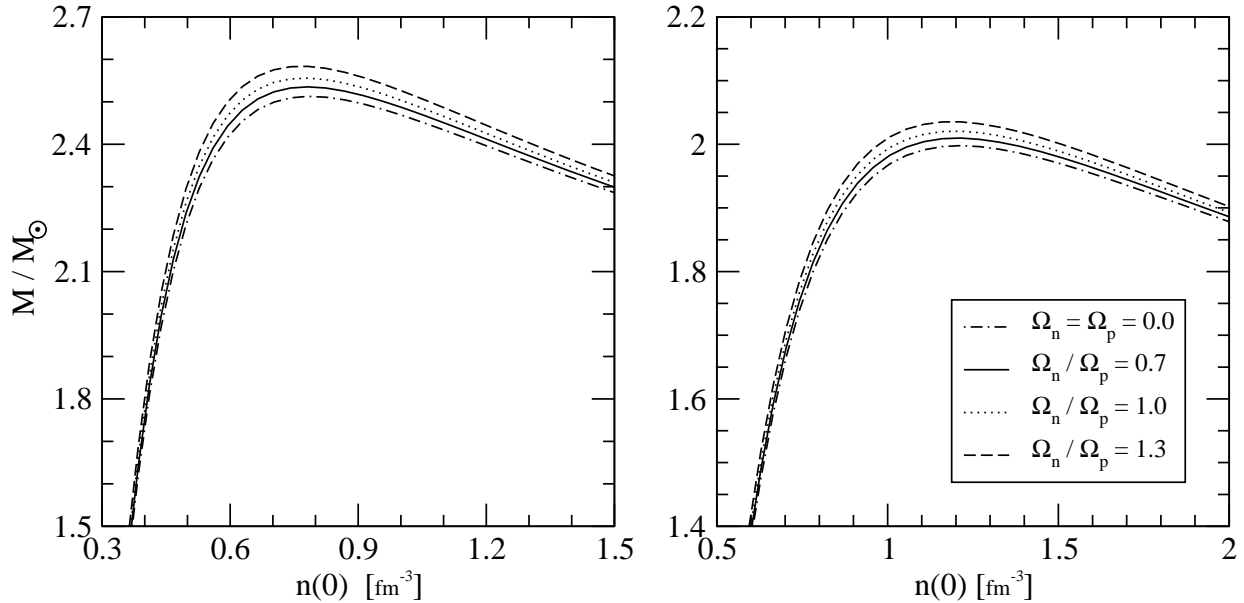}
\caption{The total mass as a function of central neutron number density 
for Models I (left panel) and II (right panel) of Table~\ref{table1} 
(using $\Omega_\p = 3900~{\rm rad/s}$). \ The relative rotation rates 
are also varied, i.e.~$\Omega_\n/\Omega_\p = 0, 0.7, 1.0, 1.3$.}
\label{masses}
\end{figure}

In Fig.~\ref{masses} we have graphed the total mass for proton rotation 
rates that equal the fastest known pulsar (i.e.~$\Omega_\p = 3900~{\rm 
rad/s}$) versus the background central neutron number density. \ The 
left panel is for Model I of Table \ref{table1} and the right panel is 
for Model II. \ In both plots the relative rotation rate $\Omega_\n / 
\Omega_\p$ of the neutrons with respect to the protons is also varied. \ 
Generally speaking, stars that are spun up without changing the total 
baryon number move upward and to the left in the figure. \ We find that 
we recover the so-called ``supramassive'' configurations first discussed 
by Cook et al \cite{CST94}. \ These are stars whose central densities go 
beyond the maximum allowed for stable, non-rotating configurations and 
yet still remain on the stable (upward) branches of their particular 
mass vs central density curves. \ These configurations are stablized by 
their rotation. 

\begin{figure}[t]
\centering
\vskip 24pt
\includegraphics[height=8.0cm,clip]{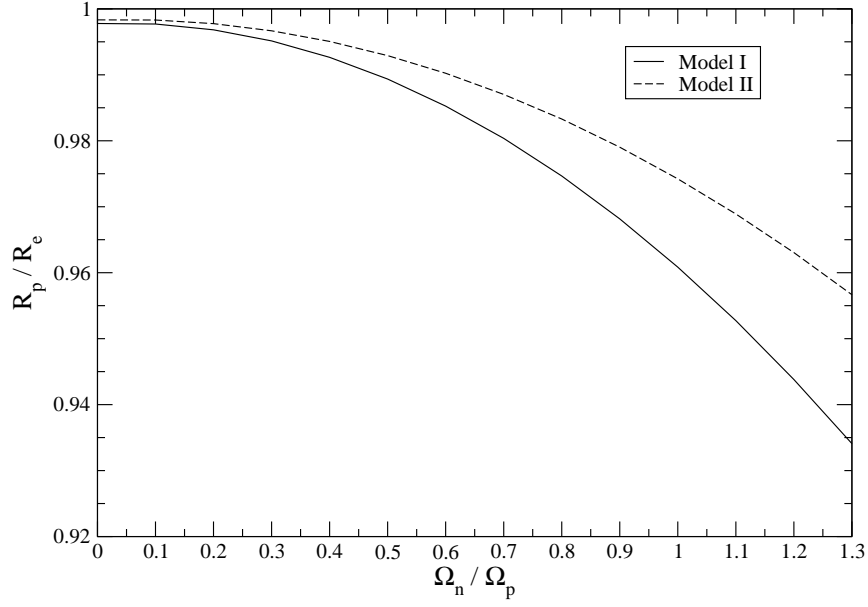}
\caption{The ratio $R_p / R_e$ as a function of the relative rotation 
rate $\Omega_\n/\Omega_\p$ for the canonical background configurations 
of Models I and II of Table~\ref{table1} and setting $\Omega_\p = 
3900~{\rm rad/s}$.}
\label{elliptic}
\end{figure}

Fig.~\ref{elliptic} gives plots of the ratio of the polar (i.e.~$R_p = R 
+ \xi(R,0)$) to equatorial (i.e.~$R_e = R + \xi(R,\pi/2)$) radii, versus 
the relative rotation rate $\Omega_\n / \Omega_\p$ for the canonical 
backgrounds given in Models I and II of Table \ref{table1} and assuming 
a proton rotation rate of the fastest known pulsar. \ Up to and 
including ${\cal O}(\Omega^2_{\n,\p})$ we have
\beq
    \frac{R_p}{R_e} = 1 + \frac{3}{2} \frac{\xi_2(R)}{R} \ .
\eeq 
We see a general overall quadratic behaviour in the relative rotation, 
which is expected for the slow rotation expansion. \ Also expected is 
the limiting value of $R_p/R_e = 1$ as $\Omega_\n/\Omega_\p \to 0$. \ 
Since the neutrons and protons have been constrained to vanish on the 
same surface, we do not find configurations like those of 
\cite{PCA02,PNC02} whereby the neutrons, say, can extend out further 
than the protons.

\begin{figure}[t]
\centering
\vskip 24pt
\includegraphics[height=8.0cm,clip]{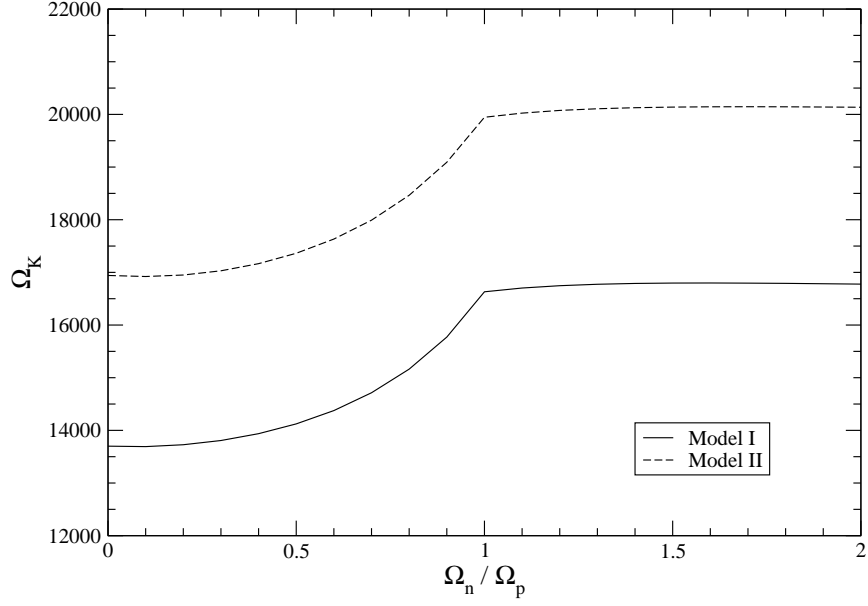}
\caption{The Kepler mass-shedding limit as a function of the relative 
rotation rate $\Omega_\n/\Omega_\p$ for the canonical background 
configurations of Models I and II of Table~\ref{table1}.}
\label{kepler}
\end{figure}

Fig.~\ref{kepler} is a graph of the Kepler (or mass-shedding) limit 
$\Omega_K$ versus the relative rotation rate for the canonical 
background configurations of Table~\ref{table1}. \ Two competing factors 
are the amount of mass of the star that is in neutrons versus the 
relative rotation rate. \ For instance, for $\Omega_\n/\Omega_\p > 1$ 
the Kepler limit is seen to be essentially flat. \ This can be easily 
understood as a result of the fact that the neutrons represent nearly 
$90\%$ of the mass of the system, and so the Kepler limit changes very 
little as the relative rotation is increased. \ Such behavior has also 
been seen in the analytical solution of Prix et al \cite{PCA02}, and the 
numerical results of Andersson and Comer \cite{AC01c}. \ The clear 
difference with the previous work is the general decrease in the Kepler 
limit as the relative rotation is decreased below one. \ Prix et al 
and Andersson and Comer found that the Kepler limit increased 
monotonically as the relative rotation was decreased. \ They explained 
this as a result of the fact that most of the mass is in the neutrons, 
and as the relative rotation is decreased, the Kepler limit is 
approaching the non-rotating value. \ As one can see in 
Fig.~\ref{kepler} eventually a minimum is reached, and beyond that the 
Kepler limit then starts to increase as the relative rotation is 
decreased. \ It just does not exhibit the monotonic increase of the 
earlier studies. \ The main differences with the two earlier studies are 
that here we use the mean field formalism for the equation of state --- 
the previous two studies used simplified polytropes --- and the 
entrainment is both relativistic and also has a non-trivial radial 
profile (cf.~Figs.~2 and 3 of \cite{CJ03}), whereas the relevant 
entrainment parameter is kept constant in Prix et al and taken to be 
zero in Andersson and Comer. 

\section{Conclusions}

The earlier work of Comer and Joynt \cite{CJ03} laid the foundations for 
a fully relativistic approach to entrainment in superfluid neutron 
stars. \ They applied their formalism to first order in the rotational 
rates, and determined the impact of a relative rotation on the 
frame-dragging and the angular momenta of the two fluids. \ We have 
expanded on this work by extending the calculations to second order in 
the rotation rates, and have determined for the first time the maximum 
mass, shape, and Kepler limit using a fully relativistic formulation for 
entrainment and allowing a relative rotation between the two fluids. 

An important extension of our work will be to include isotopic spin 
terms in the master function \cite{G97}. \ These are important for 
incorporating symmetry energy effects which tend to force baryonic 
systems to have as many protons as neutrons. \ Obviously nuclei tend to 
have equal numbers of neutrons and protons, and so this is why a 
symmetry energy term is an important addition to the equation of state. 
\ In terms of neutron stars, Prix et al \cite{PCA02} have shown clearly 
that symmetry energy impacts rotational equlibria (e.g.~ellipticity, the 
Kepler Limit, and moment of inertia) as the relative rotation is varied. 

The next application we have in mind is to study quasinormal modes on 
slowly rotating backgrounds. \ Andersson and Comer \cite{AC01b} (see 
also \cite{ACL02}) have shown that a suitably advanced gravitational 
wave detector should be able to see gravitational waves emitted during 
Vela or Crab-like glitches. \ Such observations could provide 
potentially unique information on the supranuclear equation of state and 
the parameters that govern entrainment. \ Another important application 
will be to study further the recently discovered two-stream instability 
\cite{ACP02a,PCA04}. \ It is the direct analog for superfluids of the 
instability of the same name known to exist in plasmas \cite{F63,B63}. \ 
Finally, our results should also find some application in studies of 
vortex structure in neutron stars. \ Link \cite{L03} has suggested that 
free precession in neutron stars is not compatible with the protons 
behaving as a Type II superconductor. \ This should be explored in more 
detail by allowing for the effect of entrainment.

\acknowledgments

I thank N.~Andersson for useful discussions and with providing aid with 
the code that served as the starting point for the research presented 
here. \ I also thank Bob Joynt for useful discussions and for pointing 
me in the direction of mean field theory for these entrainment 
calculations. \ The research presented here received support from NSF 
grant PHYS-0140138.
\section*{Appendix}

For convenience, we list here the various matter coefficients that are 
required as input in the field equations:
\begin{eqnarray}
\left.\A\right|_{\rm o} &=& c_{\omega}^2 + \frac{c^2_{\omega}}{5 
        \mu^2_{\rm o}} \left(2 k^2_\p \frac{\sqrt{k^2_\n + \ms0^2}} 
        {\sqrt{k^2_\p + \ms0^2}} + \frac{c^2_{\omega}}{3 \pi^2} 
        \left[\frac{k^2_\n k^3_\p}{\sqrt{k^2_\n + \ms0^2}} + \frac{k^2_\p
        k^3_\n}{\sqrt{k^2_\p + \ms0^2}}\right]\right) + \cr
        && \cr
        && \frac{3 \pi^2 k^2_{\p}}{5 \mu^2_{\rm o} k^3_\n} \frac{k^2_\n 
        + \ms0^2}{\sqrt{k^2_\p + \ms0^2}} \ , \\
        && \cr
\left.\B\right|_{\rm o} &=& \frac{3 \pi^2 \mu_{\rm o}}{k^3_\n} -
        c_{\omega}^2 \frac{k^3_\p}{k^3_\n} - \frac{c^2_{\omega} k^3_\p}
        {5 \mu^2_{\rm o} k^3_\n} \left(2 k^2_\p \frac{\sqrt{k^2_\n + 
        \ms0^2}}{\sqrt{k^2_\p + \ms0^2}} + \right.\cr
        && \cr
        && \left.\frac{c^2_{\omega}}{3 \pi^2} 
        \left[\frac{k^2_\n k^3_\p}{\sqrt{k^2_\n + \ms0^2}} + 
        \frac{k^2_\p k^3_\n}{\sqrt{k^2_\p + \ms0^2} }\right]\right) - 
        \frac{3 \pi^2 k^5_{\p}}{5 \mu^2_{\rm o} k^6_\n} \frac{k^2_\n 
           + \ms0^2}{\sqrt{k^2_\p + \ms0^2}} \ , \\
        && \cr
\left.\C\right|_{\rm o} &=& \frac{3 \pi^2 \chi_{\rm o}}{k^3_\p} -
        c_{\omega}^2 \frac{k^3_\n}{k^3_\p} - \frac{c^2_{\omega} k^3_\n}
        {5 \mu^2_{\rm o} k^3_\p} \left(2 k^2_\p \frac{\sqrt{k^2_\n + 
        \ms0^2}}{\sqrt{k^2_\p + \ms0^2}} + \right.\cr
        && \cr
        && \left.\frac{c^2_{\omega}}{3 \pi^2} 
        \left[\frac{k^2_\n k^3_\p}{\sqrt{k^2_\n + \ms0^2}} + 
        \frac{k^2_\p k^3_\n}{\sqrt{k^2_\p + \ms0^2} }\right]\right) - 
        \frac{3 \pi^2}{5 \mu^2_{\rm o} k_\p} \frac{k^2_\n + \ms0^2} 
        {\sqrt{k^2_\p + \ms0^2}} + \cr
        && \cr
        &&\frac{3 \pi^2}{k^3_\p} \sqrt{k^2_{\p} + m^2_e} \ , \\
        && \cr
\left.\a00\right|_{\rm o} &=& - \frac{\pi^4}{k^2_\n k^2_\p} 
        \frac{\partial^2 \lambda_0}{\partial k_\p \partial k_\n} = 
        c_\omega^2 + \frac{\pi^2}{k^2_\p} \frac{\ms0 \frac{\partial \ms0}
        {\partial k_\p}}{\sqrt{k^2_\n + \ms0^2}}\ , \\
        && \cr
\left.\b00\right|_{\rm o} &=& \frac{\pi^4}{k^5_\n} \left(2 
        \frac{\partial \lambda_0}{\partial k_\n} - k_\n \frac{\partial^2 
        \lambda_0}{\partial k^2_\n}\right) = c_\omega^2 + \frac{\pi^2}
        {k^2_\n} \frac{k_\n + \ms0 \frac{\partial \ms0}{\partial k_\n}}
        {\sqrt{k^2_\n + \ms0^2}} \ , \\
        && \cr
\left.\c00\right|_{\rm o} &=& \frac{\pi^4}{k^5_\p} \left(2 \frac{\partial
        \lambda_0}{\partial k_\p} - k_\p \frac{\partial^2 \lambda_0}
        {\partial k^2_\p}\right) = c_\omega^2 + \frac{\pi^2}{k^2_\p} 
        \frac{k_\p + \ms0 \frac{\partial \ms0}{\partial k_\p}}
        {\sqrt{k^2_\p + \ms0^2}} + \frac{\pi^2}{k_\p} \frac{1}
        {\sqrt{k^2_\p + m^2_e}} \ , 
\end{eqnarray}
where
\begin{eqnarray}
     \frac{\partial \ms0}{\partial k_\n} &=& - \frac{c_\sigma^2}{\pi^2} 
          \frac{\ms0 k^2_\n}{\sqrt{k^2_\n + \ms0^2}} \cr
          && \cr
          && \left(\frac{3 m - 2 \ms0}{\ms0} - \frac{c_\sigma^2}{\pi^2} 
          \left[\frac{k^3_\n}{\sqrt{k^2_\n + \ms0^2}} + \frac{k^3_\p}
          {\sqrt{k^2_\p + \ms0^2}}\right]\right)^{- 1} \ , \\
          && \cr
     \frac{\partial \ms0}{\partial k_\p} &=& - \frac{c_\sigma^2}{\pi^2} 
          \frac{\ms0 k^2_\p}{\sqrt{k^2_\p + \ms0^2}} \cr
          && \cr
          && \left(\frac{3 m - 2 \ms0}{\ms0} - \frac{c_\sigma^2}{\pi^2} 
          \left[\frac{k^3_\n}{\sqrt{k^2_\n + \ms0^2}} + \frac{k^3_\p}
          {\sqrt{k^2_\p + \ms0^2}}\right]\right)^{- 1} \ .
\end{eqnarray}

The two functions $\mu_{\rm o}$ and $\chi_{\rm o}$ are the background 
values for the two chemical potentials:
\beq
    \mu_{\rm o} = \frac{c_\omega^2}{3 \pi^2} \left(k_\n^3 + k_\p^3\right)
                  + \sqrt{k^2_\n + \ms0^2} 
                  \quad , \quad
    \chi_{\rm o} = \frac{c_\omega^2}{3 \pi^2} \left(k_\n^3 + k_\p^3
                   \right) + \sqrt{k^2_\p + \ms0^2} \ .
\eeq
Chemical equilibrium for the background means $\mu_{\rm o} = 
\chi_{\rm o} + \sqrt{k^2_\p + m^2_e}$ must be imposed. \ Given a value 
for $k_\n(0)$ then $k_\p(0)$ can be determined. \ Using these then the 
initial value for the Dirac effective mass can be obtained from 
Eq.~(\ref{diracmass}) so that Eq.~(\ref{msprime}) can be integrated.


\begin{thebibliography}{99}

\bibitem{RM69}V.~Radhakrishnan and R.N.~Manchester, Nature (London)
{\bf 244}, 228 (1969).
\bibitem{L93}A.G.~Lyne, in {\it Pulsars as Physics
Laboratories}, eds.~R.D.~Blandford, A.~Hewish, A.G.~Lyne, and
L.~Mestel, (Oxford University Press Inc., New York, 1993).
\bibitem{RD69}P.E.~Reichley and G.S.~Downs, Nature {\bf 222},
229 (1969).
\bibitem{BPPR69}G.~Baym, C.~Pethick, D.~Pines, and M.~Ruderman, Nature
{\bf 224}, 872 (1969).
\bibitem{AI75}P.W.~Anderson and N.~Itoh, Nature {\bf 256}, 25 (1975).
\bibitem{AAPS84a}M.A.~Alpar, P.W.~Anderson, D.~Pines, and J.~Shaham,
Ap.~J.~{\bf 276}, 325 (1984).
\bibitem{AAPS84b}M.A.~Alpar, P.W.~Anderson, D.~Pines, and J.~Shaham,
Ap.~J.~{\bf 278}, 791 (1984).
\bibitem{CJ03}G.L.~Comer and R.~Joynt, Phys.~Rev.~D {\bf 68}, 
023002 (2003).
\bibitem{DFM02}H.~Dimmelmeier, J.A.~Font, and E.~M\"uller, 
Astron.~Astrophys.~{\bf 393}, 523 (2002). 
\bibitem{BNM01}S.W.~Bruenn, K.R.~De Nisco, and A.~Mezzacappa, 
Ap.~J.~{\bf 560}, 326 (2001).
\bibitem{W99}F.~Weber, {\em Pulsars as Astrophysical Laboratories for 
Nuclear and Particle Physics}, (Institute of Physics Publishing, Bristol 
and Philadelphia, 1999).
\bibitem{PCA02}R.~Prix, G.L.~Comer, and N.~Andersson, 
Astron.~Astrophys.~{\bf 381}, 178 (2002).
\bibitem{ACL02}N.~Andersson, G.L.~Comer, and D.~Langlois, 
Phys.~Rev.~D {\bf 66}, 104002 (2002).
\bibitem{PNC02}R.~Prix, J.~Novak, G.L.~Comer,``Stationary structure of 
relativistic superfluid neutron stars,'' in {\em Proceedings of the 
26th Spanish Relativity Meeting: Gravitation and Cosmology}, 
eds.~A.~Lobo, F.~Fayos, J.~Garriga, E.~Gazta\~naga, and E.~Verdaguer 
(University of Barcelona, 2003), pgs.~212--217.
\bibitem{W95}J.~D.~Walecka, {\em Theoretical Nuclear and Subnuclear 
Physics}, (Oxford University Press, New York, 1995).
\bibitem{G97}N.K.~Glendenning, {\em Compact Stars: Nuclear Physics, 
Particle Physics, and General Relativity}, (Springer-Verlag, New York,
1997).
\bibitem{CLL99}G.L.~Comer, D.~Langlois, and L.M.~Lin, Phys.~Rev.~D 
{\bf 60}, 104025 (1999).
\bibitem{AC01c}N.~Andersson and G.~L.~Comer, 
Class.~and Quant.~Grav.~{\bf 18}, 969 (2001).
\bibitem{AB76}A.F.~Andreev and E.P.~Bashkin, Sov.~Phys.~JETP {\bf 42}, 
164 (1976).
\bibitem{S89}J.~Sauls, in {\em Timing Neutron Stars}, 
eds.~H.~\"Ogelman and E.P.J.~van den Heuvel, (Dordrecht, Kluwer, 
1989), pp.~457-490. 
\bibitem{BJK96}M.~Borumand, R.~Joynt, and W.~Kl\'uzniak, Phys.~Rev.~C 
{\bf 54}, 2745 (1996).
\bibitem{MTW}C.W.~Misner, K.~Thorne, and J.A.~Wheeler,
{\em Gravitation} (Freeman, San Francisco, 1973).
\bibitem{C89}B.~Carter, ``Relativistic fluid dynamics'', A.~Anile and 
M.~Choquet-Bruhat, Eds., Springer-Verlag (1989).
\bibitem{CL94}G.~L.~Comer and D.~Langlois, Class.~and 
Quant.~Grav.~{\bf 11}, 709 (1994).
\bibitem{CL95}B.~Carter and D.~Langlois, Phys.~Rev.~D {\bf 51}, 5855 
(1995).
\bibitem{CL98a}B.~Carter and D.~Langlois, Nucl.~Phys.~B {\bf 454}, 
402 (1998).
\bibitem{CL98b}B.~Carter and D.~Langlois, Nucl.~Phys.~B {\bf 531}, 
478 (1998).
\bibitem{LSC98}D.~Langlois, A.~Sedrakian, and B.~Carter, 
Mon.~Not.~R.~Astron.~Soc.~{\bf 297}, 1189 (1998).
\bibitem{P00}R.~Prix, Phys.~Rev.~D {\bf 62}, 103005 (2000).
\bibitem{C02}G.L.~Comer, Found.~Phys.~{\bf 32}, 1903 (2002).
\bibitem{H67}J.B.~Hartle, Ap.~J.~{\bf 150}, 1005 (1967).
\bibitem{FIP86}J.L.~Friedman, J.R.~Ipser and L.~Parker, 
Ap.~J.~{\bf 304}, 115 (1986).
\bibitem{CST94}G.B.~Cook, S.L.~Shapiro, and S.A.~Teukolsky, 
Ap.~J.~{\bf 422}, 227 (1994). 
\bibitem{AC01b}N.~Andersson and G.~L.~Comer, Phys.~Rev.~Letts. 
{\bf 24}, 241101 (2001).
\bibitem{ACP02a}N.~Andersson, G.~L.~Comer, and R.~Prix,  
Phys.~Rev.~Letts.~{\bf 90}, 091101 (2003).
\bibitem{PCA04}R.~Prix, G.L.~Comer, and N.~Andersson, 
Mon.~Not.~R.~Astron.~Soc., in press (2004).
\bibitem{F63}D.T.~Farley, Phys.~Rev.~Lett.~{\bf 10}, 279 (1963).
\bibitem{B63}O.~Buneman, Phys.~Rev.~Lett.~{\bf 10}, 285 (1963).
\bibitem{L03}B.~Link, Phys.~Rev.~Lett.~{\bf 91}, 101101 (2003).
\end{thebibliography}
\end{document}